\documentclass{article}
\usepackage{graphicx}
\usepackage[utf8]{inputenc}
\usepackage[T1]{fontenc}
\usepackage{natbib} 
\usepackage[dvipsnames]{xcolor}
\usepackage{soul}
\usepackage{authblk}
\usepackage{hyperref}
\hypersetup{colorlinks,citecolor={blue},urlcolor={red}}  

\usepackage{url}
\usepackage{relsize}

\usepackage{amsmath}
\usepackage{amssymb}
\usepackage{txfonts}

\usepackage{natbib}
\newcommand{\F}{{\mathcal F}}

\title{Continuous gravitational waves from neutron stars: current status and prospects}

\author[1]{Magdalena Sieniawska\footnote{Correspondence: msieniawska@camk.edu.pl}}
\author[1]{Micha\l{} Bejger}

\affil[1]{Nicolaus Copernicus Astronomical Center, Polish Academy of Sciences, Bartycka 18, 00--716 Warszawa, Poland}

\date{}

\begin{document}

\maketitle

\begin{abstract}
Gravitational waves astronomy allows us to study objects and events invisible in electromagnetic waves. It  {is} crucial to validate the theories and models of the most mysterious and extreme matter in the Universe: the neutron stars. In addition to inspirals and mergers of neutrons stars, there are currently a few proposed mechanisms that can trigger radiation of long-lasting gravitational radiation  {from} neutron stars, such as e.g., elastically and/or magnetically driven deformations: mountains on the stellar surface supported by the elastic strain or magnetic field, free precession, or unstable oscillation modes (e.g., the r-modes).  {The} astrophysical motivation for continuous gravitational waves searches, current LIGO and Virgo strategies of data analysis and prospects are reviewed in this work.

\end{abstract}










\section{Introduction}
\label{sect:intro} 

Gravitational-wave (GW) astronomy has been one of the fastest-growing fields  {in astrophysics} since the first historical detection of a binary black-hole (BH) system GW150814 \citep{Abbott2016a}. In~addition to studying the nature of
 gravitation itself, it may be used to infer information about the astrophysical sources emitting the GWs. This review concentrates on a specific kind of prospective GWs: persistent  {(continuous) gravitational waves (CGWs),} emitted by neutron stars (NSs). The~ {article} is arranged as follows. Section~\ref{sect:intro} gathers introductory material: Section~\ref{sect:basicsgr} presents the basics of the GWs theory, Section~\ref{sect:history} contains a brief  {overview} of GWs detections, Section~\ref{sect:ns} describes properties of NSs and features of hitherto detected NSs-related GWs---a binary NS merger GW170817 \citep{Abbott2017d}, Section~\ref{sect:generalcgw} gathers general information about CGWs, whereas Section~\ref{sect:methods}  { is devoted to the} main data analysis methods used in CGWs searches. Following sections  {describe the} main CGWs emission mechanisms: elastic deformations (Section~\ref{sect:eladeform}), magnetic field (Section~\ref{sect:magdeform}), oscillations (Section~\ref{sect:oscill}), free precession (Section~\ref{sect:preces}).  {Finally, in} Section~\ref{sect:summary} contains summary and~discussion. 

\subsection{Basics of the Gravitational Radiation~Theory}
\label{sect:basicsgr}

According to the general theory of relativity~\citep{Einstein1916, Einstein1918}, GWs are perturbations in the curvature of space-time, travelling with the speed of light. To~produce waves, just as in the case of electromagnetic (EM) waves, accelerated movement of charges (masses) is needed. The~lowest allowed multipole is the quadrupole, as~the monopole is forbidden by the mass conservation and the dipole by the momentum conservation. A~non-negligible time-varying quadrupole moment may be provided by e.g.,~binary systems: BHs or NSs, rotating non-axisymmetric objects (i.e., deformed NSs) or non-spherical explosions (supernov{\ae}). According to the quadrupole formula at the lowest order \citep{Einstein1916, Einstein1918}, GW amplitude strain tensor $h_{ij}$ at position $r$ is
\begin{equation}
h_{ij} = \frac{2G}{c^4 r}\ddot{Q}^{TT}_{ij}\left(t - \frac{r}{c}\right),\quad
\text{where}\quad 
Q^{TT}_{ij}(x) = \int \rho \left(x_i x_j - \frac{1}{3}\delta_{ij}r^2\right)d^3 x,
\label{eq:quadformmom}
\end{equation}
is the mass-quadrupole moment in the transverse-traceless (TT) gauge\footnote{In Einstein's theory, for~weak gravitational fields, space-time can be described as a metric: $\tilde{g}_{ij}\approx \eta_{ij}+h_{ij}$, where $\eta_{ij}$ is Minkowski metric and $h_{ij}$ corresponds to (small) GW perturbation. In~the TT gauge the perturbation is purely spatial $h_{0i} = 0$, and~traceless $h_{i}^{i} = 0$. From~the Lorentz gauge condition one can imply that the spatial metric perturbation is transverse: $\partial_i h_{ij}=0$.}, evaluated at the retarded time $(t - r/c)$ and $\rho$ is the matter density in a volume element $d^3 x$, at~the position $x^i$; $c$ and
 $G$ is the speed of light and gravitational constant, respectively. Extension of the above $h_{ij}$ expression includes 
second-order multipole moment, called current-quadrupole moment, given by \citet{Thorne1980}. It describes the dynamics of the mass currents that can lead to GWs emission caused by e.g.,~the r-mode instability (discussed in Section~\ref{sect:oscill}).

Equation~(\ref{eq:quadformmom}) shows that an axisymmetric NS rotating along its axis will not emit GWs, because~its mass-quadrupole moment will not vary in time. The~GW luminosity $L_{GW}$ is
\begin{equation}
L_{GW} = -\frac{dE}{dt}=\frac{G}{5c^5}\langle\dddot{Q}_{ij} \dddot{Q}^{ij}\rangle 
\sim \frac{G\epsilon^2 I_{3}^2 \nu^6}{c^5}, 
\end{equation}
with $\langle...\rangle$ brackets denoting  {time averaging}, and~a dimensionless parameter $\epsilon$  {quantifying} the level of asymmetry. The~moment of inertia along the rotational axis $I_3$ scales with NS mass $M$ and radius $R$ as $I_3 \sim MR^2$; $\nu$ is NS rotational frequency. An~estimate of the GW strain amplitude is thus
\begin{equation}
h_0 \sim 10^2\frac{G\epsilon I_3 \nu^2}{c^4 d}, 
\end{equation} 
which is inversely proportional to the distance to source $d$. Propagation of the GWs in vacuum is governed by a standard wave equation:
\begin{equation}
\left(\frac{\partial ^2}{\partial t^2} - \nabla ^2\right) h_{ij} = \Box h_{ij} = 0,
\end{equation}
for which the simplest solution is  {a} plane wave solution:
\begin{equation}
h_{ij}(t, \mathbf{x}) = A_{ij}\cos(\omega_{GW} t - \mathbf{k}\times \mathbf{x} + \alpha_{(i)(j)})
\end{equation}
where $\mathbf{k}$ is  {a} wave 3-vector defining the propagation direction, related to the wavelength $\lambda$ as $\lambda|\mathbf{k}| = 2\pi$, $\mathbf{x}$ is  {a} 3-vector of coordinates, $A^{ij}$ is constant amplitude and $\alpha_{(i)(j)}$ is the constant initial phase. $\omega_{GW} = 2\pi f_{GW}$ is rotational (angular) frequency, while $f_{GW}$ is   {a} frequency of the gravitational~wave.

In the TT gauge, the~above equation can be rewritten in  {the} following form that depends on two independent polarisations of the GW: plus `+' and cross `$\times$' (see explanation in \citealt{Jaranowski2009}):
\begin{equation}
h_{ij}^{TT}(t, \mathbf{x}) = h_{ij}^{TT}(t-z/c) = \begin{pmatrix} 0 & 0 & 0 & 0\\ 0 & h_{+}(t, \mathbf{x}) & h_{\times}(t, \mathbf{x}) & 0\\0 & h_{\times}(t, \mathbf{x}) & -h_{+}(t, \mathbf{x}) & 0\\ 0 & 0 & 0 & 0\end{pmatrix}
\end{equation}

Here $h_{+}$ and $h_{\times}$ are the polarisations of  {a} plane wave moving in the $+z$ direction:
\begin{equation}
h_{+}(t, \mathbf{x}) = A_{+}\cos \left[ \omega_{GW} \left(t - \frac{z}{c}\right) + \alpha_{+}\right],\quad 
h_{\times}(t, \mathbf{x}) = A_{\times}\cos \left[ \omega_{GW} \left(t - \frac{z}{c}\right) + \alpha_{\times}\right]. 
\label{eq:amppolar}
\end{equation}

Full derivation can be found e.g.,~in \citet{Misner1973,Bonazzola1996,Jaranowski2009,Prix2009}. 

The response of the GW detector $h(t)$ to the wave described with above formul{\ae} can be expressed as~\citep{Schutz1987,Thorne1987,Jaranowski1994,Jaranowski2009}:
\begin{equation}
h(t) = F_{\times}(t)h_{\times}(t) + F_{+}(t)h_{+}(t)
\label{eq:timeh}
\end{equation}
where $F_{+}$ and $F_{\times}$ are the detector's antenna-pattern (beam-pattern) functions that describe its sensitivity to the wave polarisation $h_{+}$ and $h_{\times}$~\citep{Zimmermann1979, Bonazzola1996, Jaranowski1998}:
\begin{eqnarray}
F_{+}(t) = \sin\zeta\Big( a(t)\cos(2\psi) + b(t)\sin(2\psi)\Big), \\
F_{\times}(t) = \sin\zeta\Big( b(t)\cos(2\psi) - a(t)\sin(2\psi)\Big).
\label{eq:timeantenna}
\end{eqnarray}

Note that $h_{+}$ and $h_{\times}$ depend on the mechanism of gravitational radiation and will be discussed later in this review, while $F_{+}$ and $F_{\times}$ are periodic functions with  {the} period equal to one sidereal day (for ground-based detectors such as the LIGO or Virgo), due to the rotation of the Earth. Additionally, $F_{+}$ and $F_{\times}$ depend on the  {wave polarisation angle} $\psi$, the~angle between detector's arms $\zeta$ (usually $\zeta = \pi/2$), and~two amplitude modulation functions $a(t)$ and $b(t)$, which depend on the location and orientation of the detectors on Earth and the position of the GWs source on the sky (full representation of $a(t)$ and $b(t)$ can be found e.g.,~in \citealt{Jaranowski2009}, Appendix C). Order of magnitude of frequency variations due to the daily and annual motion of the Earth is thus $10^{-6}$ Hz and $10^{-4}$ Hz, respectively. 

\subsection{Brief History of Gravitational Waves~Detections}
\label{sect:history}

First indirect evidence of the existence of GWs was deduced from the observations of stars: binary systems containing white dwarfs~\citep{Paczynski1967}, and~later neutron stars, most notably the Hulse-Taylor pulsar (catalogue number: PSR B1913+16,~\citealt{Hulse1975}). The~loss of orbital energy manifests itself as the shrinkage of the orbit and in  {the} result as a  {drop} of  {the} orbital period~\citep{Peters1963}:
\begin{equation}
\dot{P}_{GR} = -\frac{192\pi G^{5/3}}{5c^5}\left(\frac{P}{2\pi}\right)^{-5/3} (1-e^2)^{-7/2} \left(1 +\frac{73}{24}e^2 + \frac{37}{96}e^4 \right)M_1 M_2 (M_1 + M_2)^{-1/3},
\end{equation}
where $P$--- orbital period, $e$---orbital eccentricity, $M_1$ and $M_2$---masses of the components. In~30 years of observations (1975-2005) observed orbital decay was consistent  {up to} $0.13 \pm 0.21\%$ level with the theoretical prediction for the emission of GWs~\citep{Weisberg2005}. The~observed orbital decay~\citep{Taylor1982, Taylor1989, Weisberg2010} is in an excellent agreement with the expected loss of energy due to the GW radiation as described by  {the} general~relativity.

In 2015, the~real GW astronomy began with the first direct detection of the binary BH coalescence, GW150914~\citep{Abbott2016a}. The~signal was registered by two LIGO interferometers~\citep{Aasi2015b} and analysed jointly by the LIGO and Virgo Collaborations (LVC). Thus far, nine more binary BH mergers~\citep{Abbott2016b, Abbott2017a, Abbott2017b, Abbott2017c, Abbott2018b} were reported. Virgo Collaboration operates a third detector of the global network~\citep{Acernese2014}. A~second breakthrough came with the LVC observation of the binary NS merger~\citep{Abbott2017d}.  {A network of three GW detectors} cooperated with many EM observatories, performing first observations of GWs and a broad spectrum of EM waves from the same source~\citep{Abbott2017e, Abbott2017f}. These unique, multi-messenger observations allowed for the first `standard siren' measurement of  {a} Hubble constant~\citep{Abbott2017g}, measurement of the {GW propagation speed}~\citep{Abbott2017d}, discovery and study of the closest kilonova event~\citep{Abbott2017h}, estimation of the progenitor properties~\citep{Abbott2017d, Abbott2017i, Abbott2019a} and study of the post-merger remnant~\citep{Abbott2017j, Abbott2019e}.

Recently, LVC published the first catalogue with all the previous detections and source parameters'~estimates\footnote{\url{https://www.gw-openscience.org}.}. 

\begin{figure}
	\centering
	\includegraphics[width=\textwidth]{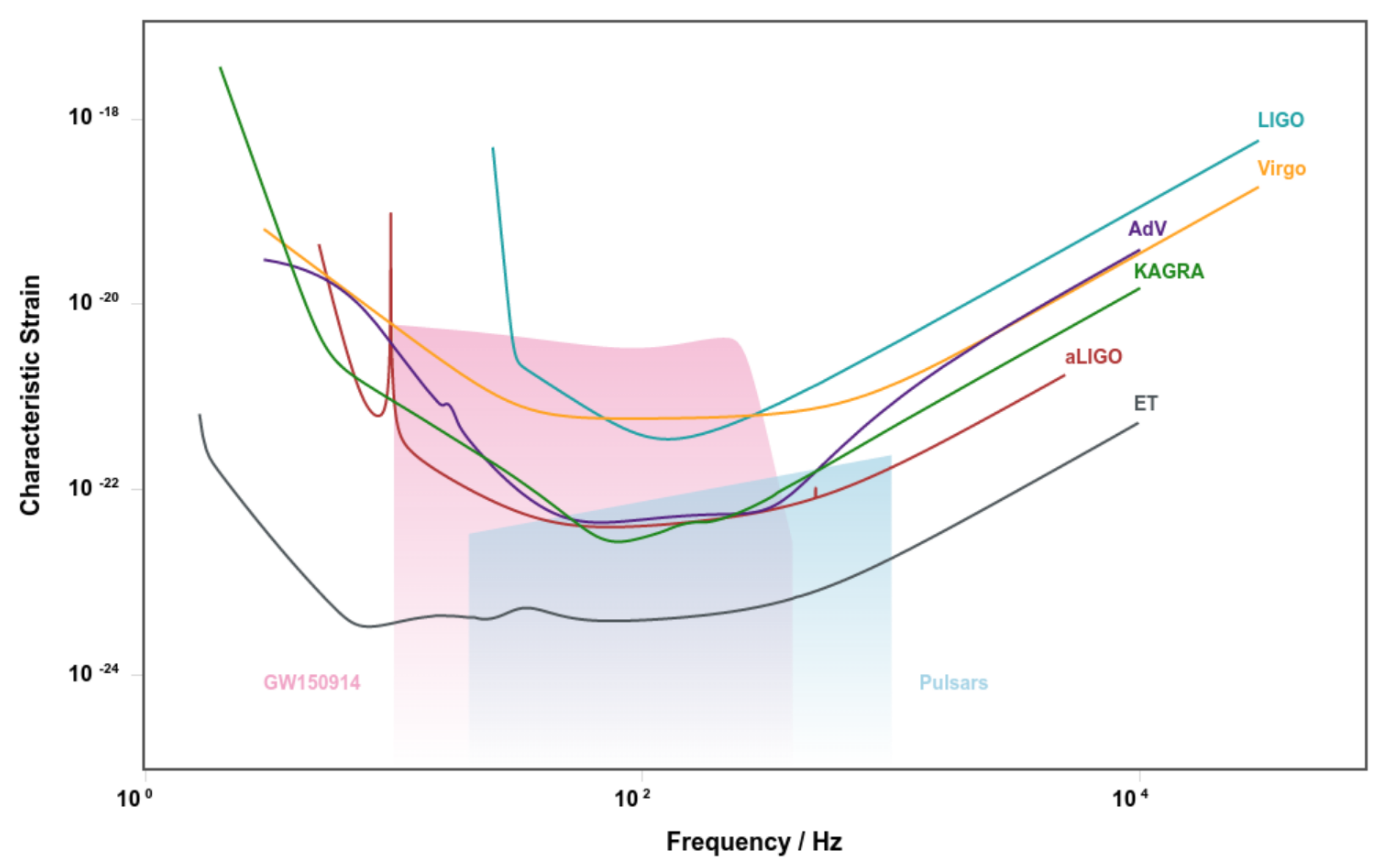}
    \caption{Comparison of the characteristic strains as a function of GW frequency for existing and planned ground-based detectors (for details see~\citealt{MooreCB2015}; plot generated with an interactive tool at {\url{http://gwplotter.com}}). The~expected CGWs amplitudes ranges of pulsars are marked with the pastel-blue region. For~comparison, the~GW150914 signal strain is represented by the pink~outline.}
    \label{fig:sensitivity}
\end{figure}

 {At the time of writing,} second-generation of interferometers: the Advanced LIGO (aLIGO,~\citealt{Harry2010}) and Advanced Virgo (AdV,~\citealt{Acernese2014})  {- are collecting data}. `Second-generation' refers to  {the strongly} improved versions of the initial, first-generation detectors. First-generation observatories were the following GW interferometers: TAMA (300-m arms) near Tokyo, Japan~\citep{Takahashi2004}, GEO600 (600-m arms) near Hannover, Germany~\citep{Willke2002}, Virgo (3-km arms) near Pisa, Italy~\citep{Freise2005}, and~LIGO (two instruments with 4-km arms each) in Hanford and Livingston, US~\citep{Abramovici1992}. Soon, next detectors will join the global GW network: the Japanese collaboration that built TAMA is now testing the 2nd-generation underground and cryogenic detector KAGRA---KAmioka GRAvitational wave telescope~\citep{Akatsu2017} and the third LIGO interferometer will be placed in India and operated by the Indian Initiative in Gravitational-wave Observations (IndIGO) in near future~\citep{Unnikrishnan2013}. Next step will be to design, build and operate third-generation detectors, with~ {the} planned sensitivity an order of magnitude better than  {the} second-generation detectors. A~European consortium is in the design stage of the Einstein Telescope (ET)---the underground trio of triangular interferometers, each 10 km long~\citep{Sathyaprakash2012}. Comparison of the current and future ground-based detectors characteristic strains in the context of CGWs sensitivities (denoted as 'Pulsars' on the plot) is shown in Figure~\ref{fig:sensitivity}.

Underground interferometers will probe GW frequencies down to ${\sim}$1 Hz, but~to reach even lower frequencies (astrophysically interesting from the point of view of e.g.,~extreme mass ratio inspirals, heavy binary systems inspiral, or~primordial GWs fluctuations from the early Universe), space-based interferometry is required. A~visionary project LISA (Laser Interferometer Space Antenna), led by European Space Agency includes a triangular configuration (each arm $2.5\times 10^6$ km long) of three satellites that will be placed in  {a} solar orbit at the same distance from the Sun as the Earth~\citep{Amaro2017}. The~mean linear distance between the formation and the Earth will be $5 \times 10^7$ km.

Currently, ongoing upgrades of the existing detectors, new detectors in the network, and~improvements of the data analysis methods will lead to an increase in sensitivity and  better quality of the detector data and,  {as a} result, to~the detections of the weaker signals.  {GW sources may be divided in to four categories, depending on the duration and strength of the signal, as~shown in} Table~\ref{tab:gwsources}: continuous gravitational waves (CGWs), subject to this review; stochastic background GW which is  {a} mixture of a large number of independent sources; inspirals and mergers of binary systems; burst sources e.g.,~supernov{\ae} explosions or magnetar flares. Search strategies for each type depend on the duration of GWs emission, knowledge of the signal model, characteristic amplitudes of the GWs, and~the computational~resources. 

\begin{table}
\begin{center}
\begin{tabular}{ | c || c | c | } 
\hline
 \quad &  Known waveform &  Unknown waveform \\ 
\hline
\hline
 Long-lived & Rotating neutron stars & Stochastic background  \\ 
 (continuous) & $h_0\sim 10^{-25}$ & $h_0\sim 10^{-28}$ \\
\hline
 Short-lived  & Compact binaries coalescences & Supernov{\ae}  \\
 ( $T\sim 0.1$ s) & $h_0\sim 10^{-21}$ & $h_0\sim 10^{-21}$\\
\hline

\end{tabular}
\caption{General taxonomy of GW sources and their expected GW strain order-of-magnitude amplitude $h_0$.}
\label{tab:gw_sources}
\end{center}
\end{table}

So far only compact objects' inspirals were detected in the LIGO-Virgo observational runs. It is expected that in the next LIGO and Virgo observing seasons, more subtle signals, such as CGWs or a stochastic background, will be detected. Steady improvement of the search methods and sensitivity of the detectors was demonstrated in the past: searches for CGWs~\citep{Abbott2017k, Abbott2018c, Abbott2019b, Abbott2019c}, stochastic background \citep{Abbott2017l, Abbott2018d} and burst signals~\citep{Abbott2017m, Abbott2019d}. Even though no significant detections were claimed, astrophysically interesting upper limits were determined. They will be discussed in the next~sections.

\subsection{Properties of Neutron~Stars}
\label{sect:ns}

 {Pioneering hypothesis of existence of the dense stars, that look like giant atomic nuclei, was given by~\citet{Landau1932}, even before the discovery of  {a} neutron\footnote{For the chronology of the events see~\citet{Yakovlev2013}.}}. Existence of the NSs, as~remnants  {of the} supernov{\ae} explosions, was proposed by~\citet{Baade1934}, just after the breakthrough discovery of  {a} neutron by~\citet{Chadwick1932}. This~hypothesis waited more than 30 years to be confirmed, when Jocelyn Bell Burnell  {first discovered a} `rapidly pulsating radio source'~\citep{Hewish1968a,Hewish1968b},  {which was} interpreted  {subsequently} as a fast-spinning NS that emits a beam of electromagnetic radiation  {similarly to} a lighthouse. Since then, the~nature of NSs is still a mystery. The~extremely large densities and pressures present inside NSs cannot be reproduced and tested in terrestrial laboratories. We also do not have a credible theoretical description of how matter behaves above the saturation density of nuclear matter $n_0=0.16$ fm$^{-3}$ (density at which energy per nucleon of infinite, symmetric nuclear matter has a minimum). 

\begin{figure}
	\centering
	\includegraphics[width=0.98\textwidth]{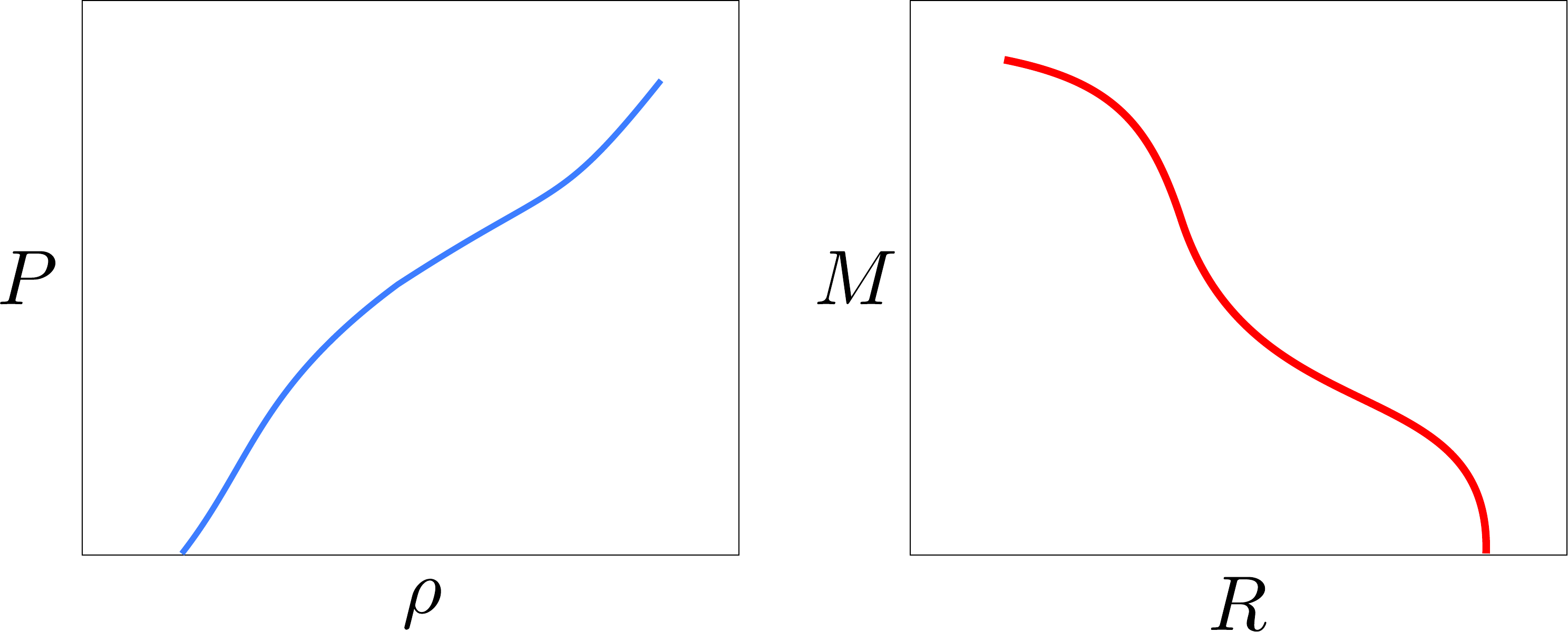}
    \caption{Schematic representation of bijective functions: measurable $M(R)$ relation (\textbf{right}), from~which currently unknown $P(\rho)$ relation (\textbf{left}), so-called equation of state, can be~established.}  
    \label{fig:eosmr}
\end{figure}

By measurements of the NSs masses $M$ and radii $R$ one can, in~principle, determine properties of  {the} matter inside NSs, such as the relation between its pressure $P$ and density $\rho$ (the currently unknown equation of state of dense matter, EOS; for a textbook introduction to the subject, see~\citealt{HPY2007book}). Bijective functions $P(\rho)$ and $M(R)$ are schematically represented on Fig.~\ref{fig:eosmr}.

Putting constraints on EOS from observations can be done  {using} the fact that $M(R)$ is  {a} bijective function to $P(\rho)$, typically by solving hydrostatic equilibrium for a spherically symmetric distribution of mass, named Tolman-Oppenheimer-Volkoff (TOV) equation~\citep{Tolman1939,OppenheimerV1939}:
\begin{equation} 
\frac{dP(r)}{dr}=-\frac{G}{r^2}\left(\rho(r)+\frac{P(r)}{c^2}\right)
\left(M(r)+\frac{4\pi r^3 P(r)}{c^2}\right)  \left(1-\frac{2GM(r)}{c^2r}\right)^{-1},
\label{eq:tov-dpdr}
\end{equation} 
 {constrained by} the equation for the total gravitational mass inside the radius $r$:
\begin{equation} 
\frac{dM(r)}{dr}=4 \pi \rho(r) r^2. 
\label{eq:tov-dmdr}
\end{equation}
 
Descriptions of the various methods used in research on EOS determination can be found in e.g.,~the review by ~\citet{Ozel2016}. From~the EM observations we know that any realistic EOS should support NSs with masses around 2 solar masses~\citep{Demorest2010, Antoniadis2013, Fonseca2016}. Uncertainties in the NSs radii and crust properties reflecting our limited knowledge of the EOS are discussed e.g.,~by~\citet{Fortin2016}. 

Although this review focuses on  {the} long-duration CGWs, for~the sake of completeness, in~the remainder of this section, we will provide a brief description of the tidal deformability effect imprinted on the GWs emitted during the last orbits of  {a} binary NS coalescence. So~far, one measurement of this kind---the GW170817 event---was successfully performed and published~\citep{Abbott2017d, Abbott2018f, Abbott2019a}, whereas a second event, dubbed S190425z, was recently reported by the LVC via the public database service~\citep{S190425z}. For~the recent theoretical studies concerning the interpretation of the GW170817 tidal deformability measurement and its relation to the dense-matter EOS, see \citet{Annala2018, Burgio2018,De2018, Fattoyev2018, Lim2018, Malik2018, Most2018, Paschalidis2018, Raithel2018, Han2019, Montana2019, Sieniawska2019a}.

In  {a} binary system,  {a} quadrupole moment of each NS is induced by the companion NS, due to the presence of  {the} external tidal field $\varepsilon_{ij}$~\citep{Misner1973,Hinderer2008}:
\begin{equation}
Q_{ij} = -\lambda_{td} \varepsilon_{ij},
\end{equation}
where the proportionality factor $\lambda_{td}$ is called the tidal deformability parameter, expressed in the lowest order approximation as
\begin{equation}
\lambda_{td} = \frac{2}{3}R^5k_2. 
\label{eq:lambdatd} 
\end{equation} 

The parameter $k_2$ is the quadrupole ($l=2$) tidal Love number~\citep{Love1911, FlanaganH2008}:
\begin{align} 
k_2 &= \frac{8}{5}x^5 (1-2x)^2 \bigl( 2-y + 2x(y-1)\bigr)   
\Bigl( 2x \bigl( 6 -3y +3x(5y-8)\bigr) \Bigr. \nonumber \\  
&+ 4x^3 \left( 13 -11y + x(3y -2) + 2x^2(1+y) \right) \nonumber \\  
&\left.+\ 3(1-2x)^2 \bigl( 2 - y + 2x(y-1)\bigr) \ln(1-2x) \right)^{-1},  
\end{align} 
with the compactness of object $x=GM/Rc^2$ ($M$ denoting the gravitational mass, $R$---radius), and~$y$---{a} solution of
\begin{align}
\frac{dy}{dr} &= -\frac{y^2}{r}-\frac{1+4\pi G r^2/c^2(P/c^2-\rho)}{(r-2GM(r)/c^2)}y  \nonumber \\
&+ \left(\frac{2G/c^2(M(r) + 4\pi r^3 P/c^2)}{\sqrt{r}(r-2GM(r)/c^2)}\right)^2 + \frac{6}{r-2GM(r)/c^2} \nonumber \\ 
&- \frac{4\pi G r^2/c^2}{r-2GM(r)/c^2}\left(5\rho+ 9P/ c^2
+\frac{\left( \rho   + P/c^2\right)^2 c^2}{ \rho dP/d\rho}\right),
\end{align}
evaluated at the stellar surface $r=R$~\citep{FlanaganH2008,VanOeverenF2017}. For~convenience, a~dimensionless value of the tidal deformability $\Lambda$ is often defined as
\begin{equation}
  \Lambda = \lambda_{td}\left( GM/c^2\right)^{-5}. 
  \label{eq:lambda} 
\end{equation} 

Tidal contribution to the GW signal are extracted from the last stages of the inspiral phase. What~is actually measured is, in~the first approximation, the~{\it effective} tidal deformability $\tilde{\Lambda}$, a~mass-weighted combination of individual dimensionless tidal deformabilities $\Lambda_1$, $\Lambda_2$, defined as
\begin{equation} 
  \tilde{\Lambda} = \frac{16}{13} 
  \frac{\left(M_1 + 12 M_2\right) M_1^4 \Lambda_1 + 
        \left(M_2 + 12 M_1\right) M_2^4 \Lambda_2} 
  {\left(M_1 + M_2\right)^5},   
  \label{eq:lambdatilde} 
\end{equation} 
with $M_1$, $M_2$ denoting the component gravitational masses\footnote{Note that the most precisely measured quantity from an inspiral phase is a so-called chirp mass: $\mathcal{M} = \frac{(M_1 M_2)^{3/5}}{(M_1+M_2)^{1/5}}$. Using Newton laws of motion, Newton universal
law of gravitation, and~Einstein quadrupole formula, one can see that $\mathcal{M}$ depends only on GW frequency $f_{GW}$ and its derivative $\dot{f}_{GW}$---quantities determined directly from the observational data: $\mathcal{M}=\frac{c^3}{G}\left[\left( \frac{5}{96}\right)^3 \pi^{-8} f_{GW}^{-11} \dot{f}_{GW}^{3} \right]^{1/5} $. Information about the individual masses is taken from the waveforms filtering, including post-Newtonian expansion. That is why $\mathcal{M}$ determination has the smallest errors, while $M_1, M_2$ estimations are model-dependent and generate relatively big errors,  {e.g., for~the GW170817 event individual masses (for the low-spin priors) were estimated as: $M_1\in(1.36, 1.60)$ M$_{\odot}$ and $M_2 \in (1.16, 1.36)$ M$_{\odot}$, while chirp mass $\mathcal{M}=1.186\pm 0.001$ M$_{\odot}$~\citep{Abbott2018f, Abbott2019a}.}}. In~principle, by~solving  {the} above equations, one can relate  {the} measurable value of $\tilde{\Lambda}$  {to} the EOS parameters $P$ and $\rho$.

\subsection{General Information about Continuous Gravitational~Waves}
\label{sect:generalcgw}

According to the ATNF (Australia Telescope National Facility) Pulsar Database\footnote{The ATNF Pulsar Database website: \url{http://www.atnf.csiro.au/people/pulsar/psrcat/}.}~\citep{Manchester2005}, more than 2700 pulsars are known.  {Assuming that CGW signal primarily corresponds to twice the NS spin frequency (in case of elastic deformations, see Section~\ref{sect:eladeform}) or close to $4/3$ of the spin frequency (Newtonian approximation in case of r-modes, see Section~\ref{sect:oscill}), around 300 pulsars from ATNF Database are} in the LIGO and Virgo detectors' frequency range  {which is the most sensitive around  50--800 Hz, as~shown in Figure~\ref{fig:sensitivity}}. Around 200 of these pulsars have precise ephemerides, as~well as, measured $\nu$ and $\dot{\nu}$. We also know from evolutionary arguments that the Galaxy contains ${\sim}10^8$--$10^9$ NSs~\citep{Narayan1987, Camenzind2005}, of~which ${\sim}$160,000 are isolated~\citep{Lorimer2005}. For~recent results concerning the population synthesis of isolated radio pulsars in the Galaxy see~\citet{Popov2007, Lorimer2011, Cieslar2018}.

As was observed by~\citet{Patruno2017}, observed limit of NSs spins is currently $\nu_{max} \approx 700$ Hz. Two possibilities were considered in~\citet{Haskell2018a}: (a) $\nu_{max}$ corresponds to  {a} maximal allowed spin, above~which  {the} centrifugal forces causes mass shedding and destroy the star (also called the Keplerian frequency). As~was shown in~\citet{Haskell2018a}, $\nu_{max}$ cannot be less than $\approx 1200$ Hz, and~that the observed lack of NSs spinning faster than $\approx 700$ Hz is not consistent with minimal physical assumptions on hadronic physics; (b) presence of additional spin-down torques acting on the NSs, possibly CGWs~emission.

It means that in our Galaxy numerous promising CGWs sources are hidden and awaiting for detections of their gravitational~signatures.

Rotating non-axisymmetric NSs~\citep{Ostriker1969,Melosh1969,Chau1970,Press1972,Zimmermann1978} are expected to emit CGWs due to the existence of time-varying mass-quadrupole moment. Such signals have smaller characteristic amplitudes than signals emitted from compact binary mergers, but~their overall duration is much longer. In~the case of almost-monochromatic CGWs, their integrated signal-to-noise ratio ($SNR$) increases with  {the} observation time $T$ as
\begin{equation}
SNR \propto h_0 \sqrt{\frac{T}{S}},
\label{eq:SNR}
\end{equation}
where $h_0$ is the instantaneous GW strain amplitude and $S$ is the amplitude spectral density of the detector's data signals' frequency. For~comparison, GW150914 lasted for $T \sim 0.2$ s in the sensitive part of detectors' band with the average GW amplitude $h_0{\sim}10^{-21}$, yielding $SNR{\sim}24$. For~CGWs, the~expected amplitude is a few order of magnitudes smaller, $h_0{\sim}10^{-25}$, but~the data  {collection} lasts for $T$ of the order of months or even years. This is one of the  {incentives for consideration of the} CGWs as serious candidates for the future detections~\citep{Brady1998, Jaranowski2000}. 

 {By simple manipulation with Equation~(\ref{eq:SNR}), one can roughly estimate the observational time that is needed to observe signals considered in this paper. Let us put $T \propto S (SNR/h_0)^2$. While~writing this paper, currently ongoing O3 observational run takes place and LIGO detector reaches $\sqrt{S}\sim10^{-23}$~Hz$^{-1/2}$ (see~\citealt{Abbott2019i} also for comparison of sensitivities of Virgo and KAGRA detectors, as~well as predictions for the future observing seasons). Typically, a~threshold for the CGW detection is set $SNR_{th} = 5$. For~the most promising scenario of CGW emission (triaxial ellipsoid, see Section~\ref{sect:eladeform}), $h_0\sim10^{-26}$ (see Equation~(\ref{eq:h0triaxe})). As~a result, in~order to detect such a signal, one needs $T\sim  300$ days of good quality, coherent data (note that O3 run is scheduled for 1 year, what makes the future CGW searches very promising). Analogously for the magnetic field distortions model (Equation~\eqref{eq:h0magn}), where $h_0\sim10^{-30} $, almost 80 mln years of data is needed! 
}

\subsection{Methods and Strategies of CGWs~Searches}
\label{sect:methods}

Detectability of the CGWs signals depends on the observational time (Equation~\eqref{eq:SNR}), but~also on the balance between computational cost of the accurate data analysis and computational resources. For~some isolated NSs the relevant parameters, such as sky position (e.g., right ascension $\alpha$ and declination $\delta$), rotational frequency $\nu$ ($f_{GW}$ is, in~case of an elastic deformation, a~mixture of $1\nu$ and $2\nu$, see Section~\ref{sect:eladeform}, or~$f_{GW} = 4/3 \nu$ for r-modes, see Section~\ref{sect:oscill}) and spin-down $\dot{\nu}$ are known from EM observations\footnote{Of course, the~whole picture is more complex when binary system is considered since in that case also the binary orbital parameters that additionally modulate the CGW signal have to be taken into account. In~this review we focus only on the isolated NSs.  {Leverage of searches for CGWs signals from isolated objects, in~order to identify and follow-up signals from NSs in binary systems were investigated in~\citet{Singh2019}}.}. For~these objects, dedicated targeted searches are performed, in~order to check if a CGW signal is associated with known parameters of the individual pulsar  {\citep{Nieder2019}}. Such searches are computationally easy to perform. A~slight modification to the targeted searches are so-called narrow-band searches,  
 which allow for a small mismatch between the frequency parameters known from the EM observations and the expected GW~signal. 

Another type of search strategy, in~which position of the source in the sky is assumed, but~the frequency parameters are unknown, is called a directed search. It may be applicable to e.g.,~a core-collapse supernov{\ae} remnants. Example of a young and relatively close supernova remnant, for~which spin frequency (and its derivatives) are unknown, is the Cassiopeia A remnant~\citep{Fesen2006}. As~was shown in~\citet{Wette2008}, CGW strain depends on the assumed age $a$ and distance $d$. Additional intrinsic parameters such as NS mass or its equation of state increase  {uncertainty for the GW strain of the strongest possible signal $h_{0}^{age}$,} by $50\%$:
\begin{equation}
h_{0}^{age} \approx 1.26 \times 10^{-24} \left( \frac{3.30 \textrm{ kpc}}{d} \right)  \left( \frac{300\textrm{ yr}}{a}\right)^{1/2}.
\end{equation}

Additionally, young and hot NSs may become unstable and undergo various dynamical processes (e.g., cooling processes and oscillations---see Section~\ref{sect:oscill}). Full realistic description of the CGWs emission, in~presence of multiple physical phenomena  may require inclusion of higher frequency derivatives and in general is extremely hard to model. Nevertheless,  {searches were performed for the CGWs} from known supernov{\ae} remnants~\citep{Abadie2010, Zhu2016, Ming2016, Ming2018, Ming2019, Abbott2019e}, as~well as in the direction towards the Galactic Centre region, where massive stars (progenitors of supernov{\ae} explosions)  {are} found in stellar clusters~\citep{Aasi2013, Dergachev2019c}.

\begin{figure}
	\centering
	\includegraphics[width=0.9\textwidth]{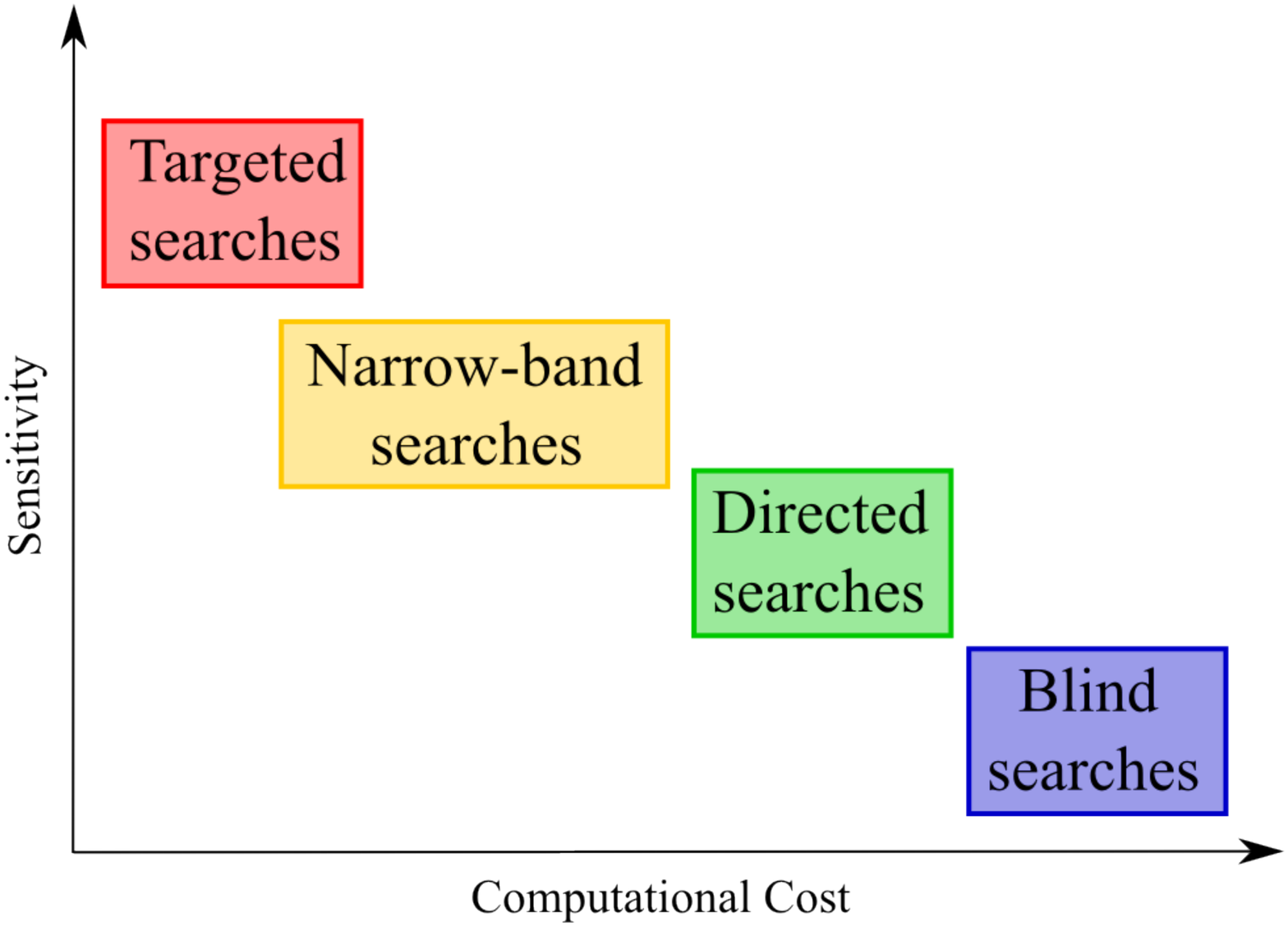}
    \caption{Schematic plot of the sensitivity versus computational cost of the different CGWs searches~strategies.}
    \label{fig:cwsearches}
\end{figure}

The most computationally intensive searches are all-sky (blind) searches, since only minimal assumptions are made to search for the signals from a priori unknown sources. Such a search requires well-optimised and robust tools, because~the less is known about the source, the~smaller sensitivity of the search can be achieved and bigger computational cost is required. All types of searches are summarised in Figure~\ref{fig:cwsearches}. Several methods (described below) were so far tested and used in blind searches on mock and real data. Each all-sky search has different advantages in the sensitivity vs. robustness against complexity of the assumed CGWs emission models, as~was compared in~\citet{Walsh2016}. 

Several search methods were developed for the CGWs signals originated in isolated NSs  {(very extensive comparison of the sensitivity of various searches can be found in~\citealt{Dreissigacker2018})}:
\begin{itemize}
\item The $\F$-statistic method introduced in~\citet{Jaranowski1998}. The~$\F$-statistic is obtained by maximizing the likelihood function with respect to four unknown parameters of the simple CGW model of rotating NSs---CGW amplitude $h_0$, initial phase of the wave $\Phi_0$, inclination angle of NS rotation axis with respect to the line of sight $\iota$, and~polarisation angle of the wave $\psi$ (which are henceforth called the extrinsic parameters). This leaves a function of only four remaining parameters: $f_{GW}$, $\dot{f}_{GW}$, $\delta$ and $ \alpha$ (called the intrinsic parameters). Thus the dimension of the parameter space that we need to search decreases from 8 to 4. To~reduce computational cost and improve method efficiency, the~$\F$-statistic can be evaluated on the 4-dimensional optimal grid of the intrinsic parameters~\citep{Pisarski2015}. As~was shown in Equation~(\ref{eq:SNR}), strength of the signal depends on the observational time: on the one hand by increasing $T$ one can expect a detection of weaker signals, on~the other hand however, analysing long-duration data requires substantial computational resources, e.g.,~for \texttt{Polgraw time-domain F-statistic} pipeline\footnote{Project repository: \url{https://github.com/mbejger/polgraw-allsky}.}, computational cost for an all-sky search scales as $\sim T^5 \log(T)$. Promising strategies to solve this problem are hierarchical semi-coherent methods, in~which data is broken into short segments. In~the first stage, each segment is analysed with the $\F$-statistic method. In~second stage, the~short time segments results are combined incoherently using a certain algorithm. Several methods were proposed for the second stage: search for coincidences among candidates from short segments~\citep{Abbott2007, Abbott2009a}, stack-slide method~\citep{Brady1998, Brady2000, Cutler2005}, PowerFlux method~\citep{Abbott2008, Abbott2009b}  {with the latest significant search sensitivity improvements for O1 data~\citep{Dergachev2019a, Dergachev2019b}}, global correlation coordinate method~\citep{Pletsch2008, Pletsch2009}, Weave method~\citep{Wette2018, Walsh2019}. Independently  {of the} details, the~main goal of the $\F$-statistic method is to find the maximum of $\F(f_{GW},\dot{f}_{GW},\delta,\alpha)$ function, and~hence the parameters associated with the signal. Several optimisation procedures (such as optimal grid-based or non-derivative algorithms) were implemented in such analyses~\citep{Astone2010a,Shaltev2013,Sieniawska2019b}. $\F$-statistic can be evaluated on the time-domain data~\citep{Jaranowski1998,Astone2010a,Aasi2014,Pisarski2015, Abbott2017k,Abbott2019b} and the frequency-domain data~\citep{Brady1998,Brady2000,Abbott2004,Abbott2007,Abbott2009a, Abbott2017p}. The~main difference between these two concepts is that in the time-domain the information is distributed across the entire data set, while the frequency-domain analysis focuses on the part of the data around the frequency at which the peak appears. The~data is initially calibrated in the time-domain and to be used by the frequency-domain methods, usually it is converted with the Fourier Transform~methods.
 
\item The Hough transform~\citep{Hough1959, Hough1962} is a  {widely used} method to detect patterns in images. It can be applied to detect the CGWs signals in specific representations of the data: on the sky~\citep{Krishnan2004}, and~in frequency-spin-down plane~\citep{Antonucci2008, Astone2014}. Both types of the Hough transform method, called \texttt{SkyHough} and \texttt{FrequencyHough}, are typically used for all-sky searches and are similar to the $\F$-statistic are matched-filtering  type methods. Due to limited computational power, they require division of data into relatively short segments. Interesting application of the Hough transform to the unknown sources searches was introduced in~\citet{Miller2018}. This \texttt{Generalised FrequencyHough} algorithm is sensitive to the braking index $n$, a~quantity that determines the frequency behaviour of an expected signal as a function of time. In~general, the~evolution (decrease) of rotational frequency is described as
\begin{equation}
\dot{\nu} = -K \nu^n,
\end{equation}
where $K$ is a positive constant. Time derivative of the above equation provides the dependency of the braking index on measurable quantities (from EM observations, e.g.,~\citealt{Espinoza2011,Hamil2015,Lasky2017,Andersson2018}), frequency and its higher derivatives:
\begin{equation}
n = \frac{\nu|\ddot{\nu}|}{\dot{\nu}^2}. 
\end{equation}
 {Value of the braking index reveals a} spin-down mechanism: $n=1$ if the spin-down is triggered by the relativistic particle wind; $n=3$, if~the spin-down is dominated by dipole radiation (as in the case of dipolar EM field); $n=5$ if it is purely quadrupolar radiation (GWs emission in General Relativity); $n=7$ if the spin-down is due to the oscillations (lowest order r-modes, see Section~\ref{sect:oscill} for further details). Some of the CGWs searches strategies assume that object is spinning-down only due to the gravitational radiation ($n=5$), as~it was mentioned with Equations~(\ref{eq:spindownlimit}) and~~\eqref{eq:epsilonlimit}, while Generalised FrequencyHough method does not assume any specific spin-down mechanism, but~allows for its~examination. 

Hough transform has also been used in the \texttt{Einstein@Home} project\footnote{Project webpage: \url{http://einstein.phys.uwm.edu}.}~\citep{Abbott2009a}, a~volunteer-based distributed computing project devoted to  {searching} for~CGWs. 

\item The 5-vector method~\citep{Astone2010b}, in~which detection of the signal is based on  {matching  a filter to} the signal $+$ and $\times$ polarization Fourier components. The~antenna response function depends on Earth sidereal angular frequency $\Omega_{\oplus}$ and results in a splitting of the signal power among five angular frequencies $\omega_{GW}$, $\omega_{GW} \pm \Omega_{\oplus}$  and $\omega_{GW} \pm 2\Omega_{\oplus}$, where $\omega_{GW}=2\pi f_{GW}$. This method is typically used for narrowband and targeted~searches.

\item The Band Sampled Data (BSD) method, is dedicated for the directed searches, or~those assuming limited sky regions, such as the Galactic Centre~\citep{Piccinni2019}.  {The application of this method results in} a gain in sensitivity at a fixed computational cost, as~well as gain in robustness with respect to source parameter uncertainties and instrumental disturbances. From~the cleaned, band-limited and down-sampled time series, collection of the overlapped short Fourier Transforms is produced. Then,  {the} inverse Fourier Transform allows removing overlap, edge and windowing effects. Demodulation of the signal from the Doppler and spin-down effects can be done e.g.,~by using heterodyne technique (see below). While in  {the} $\F$-statistic method one could manipulate with the search sensitivity by increasing the observation time, BSD method works in Fourier-domain and  {analogously} $SNR$ can be improved by increasing length of frequency bands (for comparison: bandwidth in $\F$-statistic method is typically ${\sim}0.25$ Hz and in BSD ${\sim}10$ Hz).  

\item The time-domain heterodyne method~\citep{Dupuis2005} is a targeted search which uses the EM measurements of $\nu$, $\dot{\nu}$ and $\ddot{\nu}$ (model of the phase evolution, Equation~(\ref{eq:timephase}), assumes $k=2$). The~signal depends on four unknown parameters: $h_0$, $\psi$, $\Phi_0$ and $\iota$. Due to the Earth's rotation, amplitude of the signal recorded by an interferometric detector is time-varying since the source moves through the antenna pattern (see Equations~(\ref{eq:timeh})--\eqref{eq:timeantenna}). These variations, in~the heterodyne method, are used to find characteristic frequency which is the instantaneous  {signal frequency, register at the detector}. Additionally, frequency of the signal seen in  {the} detector is affected due to the Earth motion. Second important step of the demodulation is to remove the Doppler shifts (correct signal time-of-arrival). A~targeted search is performed with a simple Bayesian parameter estimation: first the data is heterodyned with  {an} expected phase evolution and binned to short (e.g., 1 min) samples. Then, marginalisation over the unknown noise level is performed, assuming Gaussian and stationary noise over sufficiently long (e.g., of~the order of 30 minutes) periods. 95$\%$ upper limit is defined, inferred by the analysis, in~terms of a cumulative posterior, with~uniform priors on orientation and strain amplitude. At~the end the parameter estimation is done by numerical marginalisation. Effective and commonly used algorithm for the last marginalisation stage is called Markov Chain Monte Carlo~\citep{Abbott2010,Ashton2018}, in~which the parameter space is explored more efficiently and without spending much time in the areas with very low probability~densities. 

\end{itemize}

As was mentioned previously, this paper is about CGWs emission from isolated NSs. However, it is worth to mention that  NSs in binary systems are also considered as a serious candidate for sources of CGWs. NSs in binary systems have an additional modulation due to the NSs movement around the binary barycenter (at least three additional parameters). To~deal with the high computational cost, due to the bigger parameter space, search strategies usually rely on semi-coherent methods and are dedicated for known candidates (targeted/directed searches). Main proposed algorithms are: TwoSpect method~\citep{Goetz2011,Meadors2017}, CrossCorr method~\citep{Whelan2015, Meadors2018}, Viterbi/$\mathcal{J}$-statistic~\citep{Suvorova2016, Suvorova2017} or the Rome narrow-band method~\citep{Leaci2017}. Several directed searches for CGWs from NSs in known binary systems, such as Scorpius X-1, were already performed in the past~\citep{Abbott2017o, Meadors2017}. 

\section{Elastic~Deformations}
\label{sect:eladeform}

\begin{figure}
	\centering
	\includegraphics[width=0.7\textwidth]{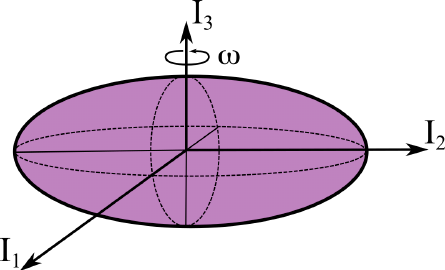}
    \caption{Simplest NS CGW model: non-axisymmetric, rotating NS (described as a triaxial ellipsoid) radiating gravitational waves at twice the spin~frequency.}
    \label{fig:triaxe}
\end{figure}

The simplest model of the NS CGWs emitter is described by  {a} rigidly rotating aligned triaxial ellipsoid, radiating purely quadrupolar waves (Figure~\ref{fig:triaxe}). As~was mentioned in Section~\ref{sect:basicsgr}, signal expected  {to be observed on the} Earth will have two polarisations $h_{\times}(t)$ and $h_{+}(t)$, which depend on the emission mechanism (Equation~\eqref{eq:timeh}):
\begin{eqnarray}
h_{+}(t) = \frac{1}{2} h_{0} \left( 1+\cos^2\iota\right)\cos(\Phi(t) + \Phi_0) \\
h_{\times}(t) = h_0\cos \iota \sin(\Phi(t) + \Phi_0),
\end{eqnarray}
where $\iota$ is  {an} inclination angle of NS rotation axis with respect to the line of sight, $\Phi(t) + \Phi_0$ is the phase of the wave ($\Phi_0$ being the initial phase), expressed as a truncated Taylor series,
\begin{equation}
\Phi(t)= \sum_{k=1}^s f_{GW}^{(k)}\frac{t^{k+1}}{(k+1)!} + \frac{\mathbf{n_0}\times \mathbf{r_d}(t)}{c}\sum_{k=1}^s  f_{GW}^{(k)}\frac{t^{k}}{(k)!},
\label{eq:timephase}
\end{equation}
where $f_{GW}^{(k)}$ is a $k$-th frequency time-derivative at the Solar System Barycentre (SSB) evaluated at $t=0$ ($f_{GW}^{(1)}=\dot{f}$, $f_{GW}^{(2)}=\ddot{f}$, ...), $\mathbf{n_0}$ is  {a} constant unit vector in the direction of the NS in the SSB reference frame (it, therefore, depends on the sky position of the source) and $\mathbf{r_d}$ is a vector joining the SSB with the~detector.

The parameter $h_0$ is a constant GW strain, which can be estimated from Einstein's quadrupole formula (Equation \eqref{eq:quadformmom}). For~the triaxial ellipsoid model it is given by the following formula:
\begin{equation}
h_{0} = \frac{4\pi^{2}G}{c^4}\frac{I_{3}f_{GW}^{2}\epsilon}{d} = 4.2 \times 10^{-26}\left(\frac{\epsilon}{10^{-6}}\right) \left(\frac{P}{10\textrm{ ms}}\right)^{-2} \left(\frac{d}{1\textrm{ kpc}}\right)^{-1},
\label{eq:h0triaxe}
\end{equation}
where the deformation (also called ellipticity) is quantified by $\epsilon = {(I_{1} - I_{2})}/{I_{3}}$. Quantities $I_1, I_2, I_3$ are the moments of inertia along three principal axes of the ellipsoid, with~$I_3$ aligned with the rotation axis (see Figure~\ref{fig:triaxe}). The~symbol $d$ denotes the distance to the source, and~$f_{GW}=2\nu$ is the GW frequency, equal to twice the rotational frequency of the star\footnote{We consider here density perturbations, which affect the spherical shape of the star $\delta\rho=Re\lbrace\delta\rho_{lm}(r)Y_{lm}(\theta, \phi)\rbrace$, where $Y_{lm}(\theta, \phi)$ denotes spherical harmonics. The~multipole moment of the perturbation along radius coordinate $r$ is $Q_{lm}=\int \delta\rho_{lm}(r) r^{l+2}dr$. Here we focus only on the lowest-order perturbation $Q_{22}$, consistent with $l=m=2$, for~which $f_{GW} = 2\nu$. Note that in this section we consider the simplest model, in~which rotational and $I_3$ axes are aligned. In~general they may be misaligned, producing additional CGW radiation at $1\nu$ frequency, whose strength depends on the angle between rotational and $I_3$ axes and is maximal when they are perpendicular~\citep{Bonazzola1996}. Such cases are consider in Section~\ref{sect:magdeform} and \ref{sect:preces}. Searches in the LVC data for the CGW radiation at both $1\nu$ and $2\nu$ were performed in the past~\citep{Abbott2019g}.}. The~NS angular frequency is given by relation $\omega=2\pi\nu={2\pi}/{P}$, where $P$ is the spin~period. 

From the right-hand side of Equation~(\ref{eq:h0triaxe}) one can notice that the ellipticity $\epsilon$ depends on the interior properties of the NS. It carries information on how easy it is to deform NS or, equivalently, about the stiffness of the matter. So far, exact values of $\epsilon$ are uncertain. The~empirical formula for the moment of inertia from~\citet{Bejger2002}, obtained for a set of realistic EOS, is
\begin{equation}
I \simeq a(x)\times MR^2,
\end{equation}
where the compactness is redefined as $x = (M/\textrm{M}_{\odot})(\textrm{km}/R)$, and~$a(x)$ functions are characteristic for the hadronic NSs and strange stars (SSs, composed of deconfined quarks, discussed below in this~section):
\begin{equation}
  a_{SS}(x) = 2(1+x)/5,\quad \quad \quad \quad 
   a_{NS}(x) =  
  \begin{cases}
    x/(0.1+2x)      & \quad \text{for } x \leqslant 0.1,\\
    2(1+5x)/9  & \quad \text{for } x>0.1.
  \end{cases}
\label{eq:aNS}
\end{equation} 

According to~\citet{Owen2005}, maximum ellipticity $\epsilon_{max}^{NS}$ depends on breaking strain of the crust $\sigma_{max}$, mass $M$ and radius $R$ in the following manner:
\begin{equation}
\epsilon_{max}^{NS} = 3.4\times10^{-7}\left(\frac{\sigma_{max}}{10^{-2}}\right)\left(\frac{1.4\textrm{  M}_{\odot}}{M}\right)^{2.2} \left(\frac{R}{10\textrm{ km}}\right)^{4.26} \left[ 1+0.7\left(\frac{M}{1.4\textrm{  M}_{\odot}}\right)\left(\frac{10\textrm{ km}}{R}\right)\right]^{-1}.
\label{eq:epsilonNS}
\end{equation}
For the typical NS parameters ($M=1.4\textrm{  M}_{\odot}$, $R=10$ km), the~maximum ellipticity can be approximated as ~\citep{Ushomirsky2000}:
\begin{equation}
\epsilon_{max}^{NS}\approx 5 \times 10^{-7} \left(\frac{\sigma_{max}}{10^{-2}}\right).
\label{eq:epsilonNSapprox}
\end{equation}

However, $\sigma_{max}$ values are also ambiguous: if the NS crust is a perfect crystal with no defects, the~$\sigma_{max} \approx0.1$~\citep{Kittel2005, Horowitz2009, Chugunov2010}. For~more amorphous materials, the~breaking strain is expected to be much smaller: $\sigma_{max}\approx10^{-5}-10^{-2}$~\citep{Ruderman1992, Kittel2005}. Nevertheless, Equation~(\ref{eq:epsilonNSapprox}) suggests that even for the extreme value of $\sigma_{max} \approx0.1$, one can expect $\epsilon_{max}^{NS}\lesssim 5\times10^{-6}$.

So far, we have assumed that NSs are made of hadronic matter,  {whose} EOS is largely unknown. Another hypothesis about the dense NS matter was proposed in the literature almost fifty years ago and is still  {waiting to be proven:~\citet{Ivanenko1965,Itoh1970,Fritzsch1973,Baym1976,Keister1976,Chapline1977,Fechner1978} considered a  possibility that in extreme conditions, e.g.,~under enormous pressure P, baryon matter is unstable. Protons and neutrons are made of two types of quarks: $u$---up and $d$
---down (proton = $uud$; neutron=$udd$) and forces between quarks are mediated by gluons. It is intuitive that the mixture of protons and neutrons can deconfine only into 2-flavour quark matter (baryons deconfine into quasi-free $u$ and $d$ quarks). Hypothetical objects made of such matter are now called quark stars (QS). }

 {However, when chemical potential grows, about half of $d$ quarks can be transformed into $s$---strange quarks. Such conversion is due to the weak-interaction process, in~contrast to the conversion of
baryons into quarks, which is a strong-interaction process (see Chapter 8 in~\citealt{HPY2007book} for the review).} Special attention will be paid to the sub-class of QS, called strange stars (SSs), in~which strange quarks---$s$ coexist with quarks $d$ and $u$~\citep{Bodmer1971, Witten1984, Haensel1986, Madsen1998}. This unique 3-flavour mixture ( {which is in weak-interaction equilibrium}) would be more stable than 2-flavour quark matter, but~also more stable than hadronic matter such as, e.g.,~iron $^{56}$Fe (which is one of the most tightly bound nuclei). For~the detailed information see  {a} review by~\citet{Weber2005}. Some authors hypothesize that SSs may be in  {a} solid state (see e.g.,~\citealt{Xu2003}),  {having also} a solid crust made of 'normal' matter. Contribution of such crust to the total ellipticity will be order of a few percent correction to that of the internal~layers.

Similarly to Equation~(\ref{eq:epsilonNS}), one can find maximal ellipticity for the solid SSs~\citep{Owen2005}:
\begin{equation}
\epsilon_{max}^{SS} = 2\times10^{-4}\left(\frac{\sigma_{max}}{10^{-2}}\right)\left(\frac{1.4\textrm{  M}_{\odot}}{M}\right)^{3} \left(\frac{R}{10\textrm{ km}}\right)^{3} \left[ 1+0.14\left(\frac{M}{1.4\textrm{  M}_{\odot}}\right)\left(\frac{10\textrm{ km}}{R}\right)\right]^{-1}.
\label{eq:epsilonNSmax}
\end{equation}

From the above, one sees that $\epsilon_{max}^{SS}$ is expected to be a few orders of magnitudes larger than $\epsilon_{max}^{NS}$. Broad study on maximal ellipticity for multiple EOS for the Newtonian and relativistic stars was performed in~\citet{Haskell2007,Mannarelli2007,Mannarelli2008,Knippel2009,Johnson2013}, 
including superconducting quark matter (SQM) for which $\epsilon_{max}^{SQM}$ can be smaller than $10^{-1}$. Observational determination of $\epsilon_{max}$ value would put strong constraints on the type and properties of the dense matter inside compact~objects. 

For more than 200 of known pulsars with known frequency and frequency derivative from EM observations, one can make the following assumption to illustrate the order of magnitude.  {Let us} assume that the observed pulsars are spinning down solely due to  {loss of the gravitational energy}. Such a hypothetical object is denoted as the {\it gravitar}~\citep{KnispelA2008}.  {For a} given detector sensitivity, an~indirect spin-down limit can be thus established~\citep{Jaranowski1998}:
\begin{equation}
h_{sd}=2.5 \times 10^{-25} \left(\frac{1 \textrm{ kpc}}{d} \right)\sqrt{\left(\frac{1\textrm{ kHz}}{f_{GW}}\right) \left(\frac{-\dot{f}_{GW}}{10^{-10}\textrm{ Hz/s}}\right)  \left(\frac{I_3}{10^{38}\textrm{ kg}\times\textrm{m}^2}\right)}.
\label{eq:spindownlimit}
\end{equation}

 {Combination of} Equations~(\ref{eq:h0triaxe}) and \eqref{eq:spindownlimit}  {results in a} corresponding spin-down limit  {for} the fiducial equatorial ellipticity:
\begin{equation}
\epsilon_{sd} = 0.237\left(\frac{I_3}{10^{38}\textrm{ kg}\times\textrm{m}^2}\right)^{-1}  \left(\frac{\textrm{Hz}}{\nu}\right)^2  \left(\frac{h_{sd}}{10^{-24}}\right)    \left(\frac{d}{1 \textrm{ kpc}} \right).
\label{eq:epsilonlimit}
\end{equation}

So far, in~the O1 and O2 LVC observing runs, no CGWs signals were detected, but~instead interesting upper limits in a broad range of frequencies for known pulsars were set~\citep{Abbott2017n, Abbott2019c}. For~example, the~currently best limits for Crab pulsar (J0534+2200) are
$h_{sd} = (1.4 \pm 0.4) \times 10^{-24}$ resulting in $\epsilon_{sd} = 7.56 \times 10^{-4}$.  {Additionally, in~the O2 data analysis, for~some of the pulsars (J1400-6325,  
J1813-1246, J1833-1034, J1952+3252, J0940-5428, J1747-2809, J1400-6325, J1833-1034, J1747-2809), the energy spin-down limits (the maximal available rotational energy loss due to CGW emission) were surpassed~\citep{Abbott2019c}. For~example, the~Crab pulsar (J0534+2200) is emitting $\sim$1\% or less energy in CGW (for more details see Table IV in~\citet{Abbott2019c}).
}

Prospective detection of CGWs, consistent with a deformed NSs, will be a breakthrough in several topics, mostly related to the properties of matter under huge pressures. It will put strong constraints on EOS, and~likely may yield a distinction between compact stars built of the hadronic and strange matter. It will also allow for testing of the outer NS layer---the crust. An~analysis of the crustal failure and its other properties may reveal not only astrophysically important information, but~also can be valuable in engineering, e.g.,~mechanics and  { material strength studies}. Finally, it allows measuring deformation itself, describe the exact NS shape and mechanisms that triggered deviations from a spherical~shape.

\section{Magnetic~Field}
\label{sect:magdeform}

In  {a} widely accepted pulsar model, inferred from the EM observations, misalignment between the global dipole magnetic field axis and the rotation axis is responsible for observed pulsations. It is likely that for some range of conditions (described in this section), an~asymmetry in the magnetic field distribution in the interior of NS may lead to the emission of~CGWs.

\begin{figure}
	\centering
	\includegraphics[width=0.55\textwidth]{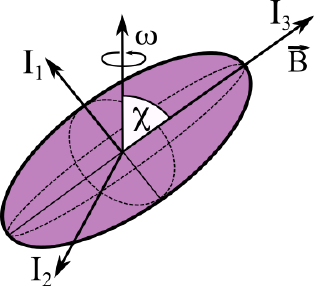}
    \caption{A model of the CGWs emission due to the magnetic deformation. The~angle $\chi$ measures the misalignment between the magnetic field axis and rotation axis. In~principle for $\chi \neq 0$, CGWs emission will be produced even if $I_1=I_2$. Note that on this schematic picture $\vec{B}$ represents the dominant toroidal field, reflecting the fact that NS shape is prolate along the magnetic~axis.}
    \label{fig:btriaxe}
\end{figure}

The idea that magnetic stresses can deform  {a} star and lead to the CGWs emission was originally proposed by~\citet{Chandrasekhar1953} and was considered later in~\citet{Bonazzola1996, Jones2002, Cutler2002, Haskell2008, Kalita2019}. The~simplest model was considered in~\citet{Galtsov1984a, Galtsov1984b}, where a NS was approximated by a rigidly rotating Newtonian incompressible fluid body, with~the uniform internal magnetic field and the dipole external magnetic field. Inclination of the magnetic dipole moment with respect to the rotation axis is given in this model by angle $\chi$. Note that for misaligned rotational and magnetic axes, time-varying quadrupole moment will also be present for $I_1=I_2$. In~such conditions, the~NS is a biaxial ellipsoid, illustrated in Figure~\ref{fig:btriaxe}. The~ellipticity is then described by the ratio $\epsilon=(I_3-I_1)/I_1$, and~the resulting CGW strain will take form~\citep{Bonazzola1996}:
\begin{equation}
h_0 = 6.48\times 10^{-30} \frac{\beta}{\sin^2(\chi)}\left(\frac{R}{10\textrm{ km}}\right)^2 \left(\frac{1\textrm{ kpc}}{d}\right) \left(\frac{1\textrm{ ms}}{P}\right)  \left(\frac{\dot{P}}{10^{-13}}\right),
\label{eq:h0magn}
\end{equation}
where coefficient $\beta$ measures efficiency of the magnetic structure in distorting the star (magnetic distortion factor). For~ {a} simplified model (incompressible fluid, uniform internal magnetic field), parameter $\beta$ equals $1/5$. Information about the induced deformation is encoded in $\beta$ factor, as~$\epsilon = \beta \mathcal{M}^2/\mathcal{M}_0^2$, where $\mathcal{M}$ is the magnetic dipole moment and $\mathcal{M}_0$ has the dimension of a magnetic dipole moment in order to make $\beta$ dimensionless. It is clear that $h_0$ grows  {when} $\chi\rightarrow 0$ (but for practical reasons $\chi$ cannot be too small because it will break the formula down), and/or for large $\beta$. As~was shown in~\citet{Bonazzola1996}, such emission occurs mostly on $f_{GW}=1\nu$. One can also notice, by~comparison of the prefactors of Equations~(\ref{eq:h0triaxe}) and \eqref{eq:h0magn} that the expected CGW strain from magnetic effects is almost four orders of magnitude smaller than in the case of triaxial ellipsoid. For~the above model, in~case of the incompressible NS with poloidal magnetic field ellipticity is expressed as~\citep{Haskell2008}:
\begin{equation}
\epsilon \approx A \left(\frac{R}{10\textrm{ km}}\right)^4 \left(\frac{M}{1.4\textrm{  M}_{\odot}}\right)^{-2}\left(\frac{\bar{B}}{10^{12}\ \textrm{G}}\right)^2,
\label{eq:epsilonBpol}
\end{equation}
where $\bar{B}$ is the volume averaged magnetic field and factor $A$ is a constant, characteristic to the model (e.g., $A=10^{-12}$ in case of a uniform density star, $A=8\times 10^{-11}$ for the polytrope $n=1$ with the purely poloidal field, $A=-5\times 10^{-12}$ for the polytrope $n=1$ with purely toroidal field, etc.)\footnote{Note that in Equation~(\ref{eq:epsilonBpol}) the function has a positive sign, which means that the poloidal magnetic field tends to distort a NS into an oblate shape. For~a toroidal field the expression changes sign, making the NS shape prolate.}. 

It was shown that purely poloidal or purely toroidal magnetic fields are unstable~\citep{Wright1973, Braithwaite2006S, Braithwaite2007}, and~that the realistic description of NS requires mixed field configurations~\citep{Braithwaite2006N, Ciolfi2009, Lander2012}, giving an estimate of ellipticity~\citep{Mastrano2011}:
\begin{equation}
\epsilon_{mixed} \approx 4.5 \times 10^{-7} \left( \frac{B_{poloidal}}{10^{14}\ \textrm{G}}^2\right) \left(1 - \frac{0.389}{\Xi}\right),
\end{equation} 
where $B_{poloidal}$ is the poloidal component of the surface magnetic field, and~$\Xi$ is the ratio of poloidal-to-total magnetic field energy: $\Xi = 0$ corresponds to purely toroidal field and $\Xi = 1$ to purely poloidal field.
NSs have magnetic fields below $10^{15}$ G~\citep{Shapiro1983}, so the effect of magnetic deformation is significantly smaller than the one discussed in Section~\ref{sect:eladeform}. NSs with magnetic fields typically of the order of $10^{15}$ G are called {\it magnetars}. Known magnetars rotate with periods of $P {\sim}$ 1--10 s~\citep{Manchester2005}, so the frequency of the emitted CGWs would be very low, in~the frequency range unavailable for Advanced LIGO, Advanced Virgo and next generation detectors, such as the Einstein Telescope~\citep{Lasky2015}. 

Ref. \citet{Cutler2002} and recent evaluations for different EOS and bulk viscosity models in~\citet{DallOsso2018} show that, while the first-year spin-down of a newborn NS is most likely dominated by EM processes, reasonable values of internal toroidal field $B_{toroidal}$ and the external dipolar field $B_{external}$ can lead to detectable GWs emission, for~an object in our Galaxy. While the centrifugal force distorts a NS into anoblate shape, the~internal toroidal magnetic field makes them prolate. $B_{toroidal}$ determines the NS shape if
\begin{equation}
B_{toroidal} \gtrsim 3.4 \times 10^{12} \left( \frac{\nu}{300\textrm{ Hz} } \right)^2\ \textrm{G}.
\end{equation}

The above formula is important not only for the newborn NSs, but~also for the NSs in binary systems, where accretion processes are active and affecting magnetic field. That may happened e.g.,~in low-mass X-ray binaries (LMXB)---systems in which mass transfer between the companion and NS is via Roche-lobe overflow. Material in accretion disc formed around NS is heated so much that it shines brightly in X-rays. NSs in LMXB initially possess $B_{external} \sim 10^{12}$--$10^{14}$ G, $B_{toroidal} \sim 10^{12}$--$10^{15}$~G and their spin and magnetic axes are nearly aligned ($\theta\approx0$), as~was studied in~\citet{Cutler2002}. Such objects spin-down electromagnetically, until~accretion process from the companion is ongoing: accretion reduces $B_{external}$ below $10^9$ G. However, in~the interior, $B_{toroidal}$ remains unchanged. Dissipation~processes dominate and at some point magnetic axis rapidly `flips' orthogonally to the rotational axis. For~the full understanding of the LMXB evolution and GWs emission, one should consider comparison between timescales of four processes: (i) spin-down due to the GWs emission:
\begin{equation}
\frac{1}{\tau_{GW}}=5.50\times 10^{-13} \left( \frac{\epsilon_B}{10^{-7}} \right)^2 \left( \frac{\nu}{1\textrm{ kHz}} \right)^4\textrm{ s}^{-1} ;
\end{equation}  

(ii) spin-down due to the EM process:
\begin{equation}
\frac{1}{\tau_{EM}}=4.88\times 10^{-16} \left( \frac{B_{external}}{10^9\ \textrm{G}} \right)^2 \left( \frac{\nu}{1\textrm{ kHz}} \right)^2 \textrm{ s}^{-1} ;
\end{equation}  

(iii) timescale on which accretion can significantly change the NS angular momentum:
\begin{equation}
\frac{1}{\tau_{ACC}}=9.30\times 10^{-16} \left( \frac{\dot{M}}{10^{-9} \textrm{ M}_{\odot}/\textrm{yr}} \right) \left( \frac{300\textrm{ Hz}}{\nu} \right) \textrm{ s}^{-1} ,
\end{equation}  
where $\dot{M}$ is the accretion rate; (iiii) dissipation timescale on which the instability acts:
\begin{equation}
\frac{1}{\tau_{DISS}}=3.00\times 10^{-8} \left( \frac{10^4}{\mathcal{Y}} \right) \left( \frac{\nu}{300\textrm{ Hz}} \right) \left( \frac{\epsilon_B}{10^{-7}} \right) \textrm{ s}^{-1} ,
\end{equation} 
where $\mathcal{Y} = \tau_{DISS}\frac{\epsilon_B}{P}$ is  {a parameter, which was found hard to estimate and $\epsilon_B$ is the deformation caused by the toroidal field and defined as:
\begin{equation}
\epsilon_B=-\frac{15}{4}\left(\frac{3GM^2}{5R}\right)^{-1}\int\frac{1}{8\pi}B_{toroidal}^2 dV
\label{eq:epsilon_B}
\end{equation}
}
 
As  {it} was shown in~\citet{DallOsso2018}, for~an optimal range of $\epsilon_B\sim(1-5)\times 10^{-3}$ and birth spin period $\lesssim$2~ms the horizon of CGWs detectability is equal  {to} 4, and~30 Mpc, for~Advanced and third generation interferometers at design sensitivity, respectively. Additionally, LMXB give opportunity to estimate maximum expected signal from accreting neutron stars, taking advantage from EM observations. If~the angular momentum lost in GWs  is recovered by accretion, then the strongest GWs emitters are those accreting at the highest rate, namely LMXB. As~ {it} was shown in~\citet{Papaloizou1978, Wagoner1984, Bildsten1998}, if~the spin-down torque from GWs emission is in equilibrium with the accretion torque, $h_0$ is directly related to the X-ray luminosity~\citep{Wagoner1984,Bildsten1998,Ushomirsky2000}:
\begin{equation}
h_0 \approx 3.5 \times 10^{-27} \left( \frac{R}{10\textrm{ km}} \right)^{3/4} \left( \frac{M}{1.4\textrm{ M}_{\odot}} \right)^{1/4} \left( \frac{300 \textrm{ Hz}}{\nu} \right)^{1/2} \left( \frac{F_x}{10^{-8} \textrm{ erg cm}^{-2}\textrm{ s}^{-1}} \right)^{1/2}, 
\label{eq:Fx-GW}
\end{equation}
where $F_x$ is the observed X-ray flux; note that here the information about the distance is encoded in $F_x$ and both---GWs and X-rays---are falling inversely proportional to $d^2$. Most of LMXB have spins in the relatively narrow range 270 Hz $\lesssim$ $\nu$ $\lesssim$ 620 Hz~\citep{Chakrabarty2003}, while the Keplerian frequency (discussed in Section~\ref{sect:generalcgw}) is typically much bigger, $\sim$1400 Hz. That leads to the conclusion that LMXB have to be equipped in an additional mechanism balancing the angular momentum transfer and preventing spin increase. According to Equation~(\ref{eq:Fx-GW}), brightest known X-rays source, Sco X-1~\citep{Giacconi1962} should also emit the strongest GW signal:
\begin{equation}
h_0^{\textrm{Sco X-1}}\approx 3\times 10^{-26} \left( \frac{540 \textrm{ Hz}}{\nu} \right)^{1/2}.
\end{equation}

Unfortunately, Sco X-1 spin frequency is not well constrained, it is nevertheless one of  {the} prime targets for LVC and its GW signal was searched for in the past~\citep{Abbott2007, Aasi2015a, Abbott2019f}. During~the latest and the most sensitive search~\citep{Abbott2019f}, no evidence for the GWs emission was found; 95\% confidence upper limits were set at $h_0^{95\%} = 3.47 \times 10^{-25}$, assuming the marginalisation over the source inclination~angle.

As was mentioned several times in this section, CGWs created by the magnetic processes typically have amplitudes smaller than the sensitivity of Advanced LIGO and Advanced Virgo. However, it may be within  {the} reach of a future improved network of detectors. Successful detection of such signals would be an amazing tool for probing the magnetic fields in NSs: their composition (poloidal, toroidal), strength and evolution. Indirectly one could test deformability and compressibility of the NSs. These studies will be also complementary to EM observations of special NSs classes: magnetars, young NSs and NSs in LMXB systems. Closer look through the GW analysis could possibly allow for testing of accretion, precession and spin evolution~processes.
\section{Oscillations}
\label{sect:oscill}

In asteroseismology, the~Rossby modes (also called r-modes,~\citealt{Rossby1939}) are a subset of inertial waves caused by the Coriolis force acting as restoring force along the surface.  {In the NSs the r-modes are triggered by the Chandrasekhar-Friedman-Schutz instability~\citep{Chandrasekhar1970,Friedman1975}.} This instability is driven by GW back-reaction---it creates and maintains hydrodynamic waves in the fluid components, which propagate in the opposite direction to that of the NS rotation, producing GWs. So far, the~EM observations of the oscillations in NSs are insufficient in the context of EOS, because~it is impossible to directly observe the modes or thermal radiation from the surface. Details of the methods and limitations of EM asteroseismological observations can be found e.g.,~in~\citet{Cunha2007}. It is expected that the CGWs observations will complement our understanding of the compact objects~oscillations.

\begin{figure}
	\centering
	\includegraphics[width=0.8\textwidth]{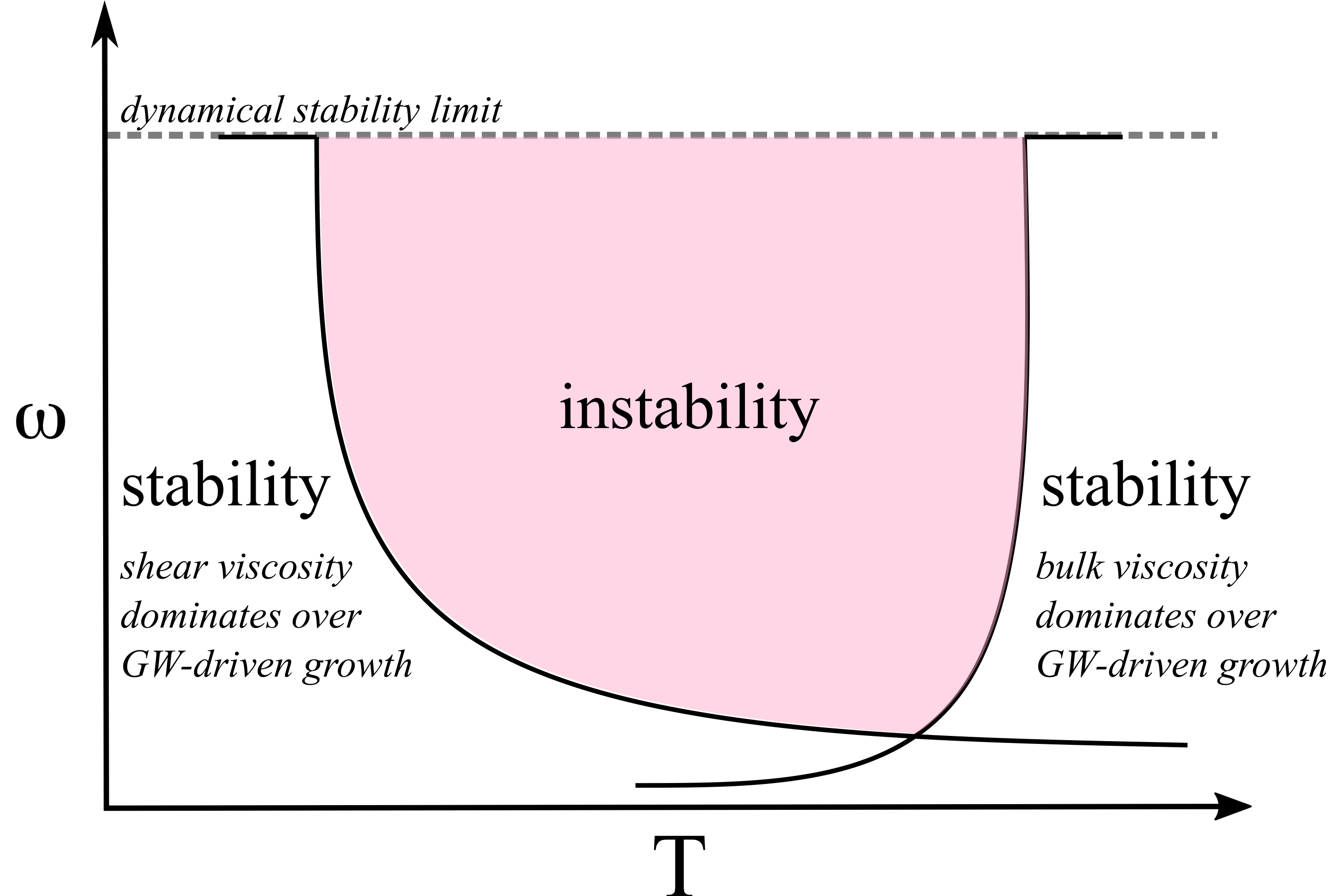}
    \caption{Schematic plot of the window in which r-modes are unstable and lead to CGWs emission.  {The shape of instability window and NSs evolutionary tracks inside and outside the window are model-dependent (including EOS-dependence, see footnote 14 and e.g.,~\citet{Haskell2012,Chugunov2017,Wang2019}). For~the more detailed review of the topic see~\citet{Kokkotas2016}.}}
    \label{fig:rmodesinsta}
\end{figure}

The r-modes were first proposed as a source of GWs from newborn NSs~\citep{Owen1998} and from accreting NSs~\citep{Bildsten1998,Andersson1999}. It was shown that the r-modes can survive only in a specific temperature window, in~which they remain unstable: at too low temperatures, dissipation due to shear viscosity damps the mode and when the matter is too hot, bulk viscosity will prevent the mode from growing, as~shown in the schematic plot in Figure~\ref{fig:rmodesinsta}. R-mode instability window is open for NSs with temperatures between $10^6$ --$10^{10}$ K~\citep{Owen1998, Bildsten1998, Bondarescu2007, Haskell2015}. This information allows testing properties of the interiors of NSs and theoretical models that describe matter interactions in such conditions.   {The simplest model\footnote{Whole energy of the mode is transferred to the NS spin-down and loss of the canonical angular momentum of the mode.} of the CGW emission, triggered by the unstable r-modes in the newborn NSs, was introduced in~\citet{Owen1998}, where authors shown that CGW amplitude in this case can be expressed by:
\begin{equation}
h_0 = 4.4\times 10^{-24}\alpha \left( \frac{\omega}{\sqrt{\pi G \bar{\rho}}} \right)^3 \left(\frac{20\textrm{ Mpc}}{d}\right),
\end{equation}
where $\alpha$ is dimensionless r-mode amplitude and $\bar{\rho}$ is the mean density.}

 {Also in the temperature range that corresponds to the r-modes instability window}, conditions to form superfluidity are fulfilled. The~superfluid component plays an important role in NSs interiors e.g.,~in attempts to explain glitches e.g.,~sudden increases of rotational frequency (these events are regularly observed, but~not explained theoretically,~\citealt{Haskell2015}). For~the NSs in LMXB systems, the~r-modes can affect the spin evolution of rapidly spinning objects~\citep{Haskell2015, Haskell2017}. This mechanism was proposed not only for young and hot NSs, but~also for the old, accreting NSs. In~the LMXB, together with matter also the angular momentum is `accreted', resulting in spinning up the object. This is thought to be the mechanism by which old, long-period NSs are recycled to millisecond periods~\citep{Alpar1982}. In~principle, long-lasting persistent accretion may spin NSs up to the Keplerian frequency (see previous Section), but~a spin frequency cut-off of about 700 Hz is observed~\citep{Chakrabarty2003}. One of the proposed mechanisms to balance the spin evolution is the GWs emission due to the r-mode instability~\citep{Bildsten1998,Andersson1999}. When the r-mode amplitude, defined in Equation~(\ref{eq:rmodeamp}), is very large, the~system will be located deeply inside the instability window, but~also the evolution will be very rapid and the system will rapidly evolve out of the window~\citep{Levin1999,Spruit1999}.  {On the} contrary, when the r-mode amplitude is low, the~evolution is slower, but~system will remain close to the instability curve~\citep{Heyl2002}. As~a consequence, in~both scenarios it is highly unlikely to observe a LMXB deeply inside the instability window. However, according to~\citet{Haskell2015}, our theoretical understanding is not fully consistent with observations: a large number of observed LMXBs is located deeply in the r-mode instability window~\citep{Haskell2012}. Mechanisms, which can potentially explain this inconsistency, are e.g.,~the presence of exotic particles in the core~\citep{Alford2012,Haskell2012}, viscous damping at the crust/core boundary layer interface~\citep{Glampedakis2006, Ho2011} or strong superfluid mutual friction~\citep{Andersson2006,Haskell2009}. An~advantage of the LMXB systems is that they are observed in the EM spectrum and some systems were tentatively interpreted as sources of r-modes recently, e.g.,~by~\citet{Andersson2018}. The~study of r-modes provides a huge opportunity to test various  models of  NSs interiors  and the physics behind their spin~evolution.

For the CGWs triggered by r-modes oscillations, the~(Newtonian) relation for the lowest order (strongest) mode\footnote{Unlike in the case of the elastic deformations, what was shown in Section~\ref{sect:eladeform}, r-modes affect equation of motion $\delta v = \alpha (r/R)^{l} R\omega Y_{lm}^{B}exp(i \omega_{r}t)$, where $\alpha$ is dimensionless amplitude and $\omega_{r} = -\frac{2m}{l(l+1)}\omega$ is the angular frequency in co-rotating frame. The~only non-trivial solution is for $l=m=2$, giving $\omega_r = -\frac{2}{3}\omega$, what can be transferred to the angular frequency $\omega_i$ in the inertial frame: $\omega_i = -\frac{2}{3}\omega + 2\omega = \frac{4}{3}\omega$. Dividing this expression by $2\pi$ gives $f_{GW} = \frac{4}{3} \nu$.} is  $f_{GW} = 4/3 \nu$. This relation is independent on EOS and it is a good approximation for the slowly rotating NSs. However, according to numerical results, inclusion of relativistic corrections may increase $f_{GW}$ by a few tens of percent~\citep{Lockitch2003,Pons2005,Idrisy2015,Jasiulek2017}. Exact value depends mostly on the object compactness $C$ (dimensionless mass-radius ratio: $C\approx 0.207\left( \frac{M}{1.4\textrm{ M}_{\odot}} \right) \left( \frac{10\textrm{ km}}{R} \right)$, for~NSs typically $0.11\lesssim C\lesssim 0.31$), but~also on rotational frequency of the object~\citep{Lindblom1999,Idrisy2015,Jasiulek2017}, including rigid versus differential rotation~\citep{Stavridis2007,Idrisy2015,Jasiulek2017}, stiffness of the crust~\citep{Bildsten2000, Levin2001} and EOS~\citep{Lockitch2003,Pons2005,Lattimer2007,Idrisy2015}.  {Additionally, as~it was shown in~\citet{Arras2003}, consideration of the non-linear coupling forces among internal modes leads to the result that r-mode signal from both newly born NSs and LMXB in the spin-down phase of Levin’s limit cycle\footnote{When viscosity is dominated by normal matter, then the NS enters into a limit cycle of spin-up by accretion and spin-down by the r-mode.}~\citep{Levin1999} will be detectable by enhanced LIGO detectors out to $\sim$100--200~kpc. Dissimilarities of the non-linear effects in supra-thermal regime for hadronic and quark EOS were studied by~\citet{Alford2010}.} Nonetheless, for~practical purposes (such as CGWs searches),  {a} Newtonian approximation is typically~used.

Potentially, the~detection of r-modes can confirm or rule out the existence of SSs, introduced and discussed in Section~\ref{sect:eladeform}. The~viscosity coefficients of NSs and SSs differ significantly and as a consequence, the~r-modes in  a SS are unstable at lower temperatures~\citep{Chatterjee2008,Alford2010,Haskell2012}. The~existence, evolution and properties of SSs is still an open question, widely discussed in the literature (see e.g.,~\citealt{Jaikumar2006, Blaschke2002}). As~shown in~\citet{Mytidis2015}, a~hypothetical r-mode detection can put constraints on the moment of inertia of the compact object and bring closer to  understanding  the~EOS.

The growth time of the instability for r-modes can be relatively short\footnote{The timescales discussed here are related to the $n=1$ polytrope  {and $l=2$}: the simplest illustrative model.  {General expressions of the timescales are~\citep{Lindblom1998}: $\frac{1}{\tau_{GW}}=-\frac{32\pi G \omega^{2l+2}}{c^{2l+3}}\frac{(l-1)^{2l}}{[(2l+1)!!]^2}\left(\frac{l+2}{l+1}\right)^{2l+2}\int\limits_0^R \rho r^{2l+2}dr$; $\frac{1}{\tau_{sv}}=(l-1)(2l+1)\int\limits_0^R \eta r^{2l}dr \left(\int\limits_0^R\rho r^{2l+2}dr\right)^{-1}$, where $\eta = 347\rho^{9/4}T^{-2}$ is the shear viscosity factor; $\frac{1}{\tau_{bv}}\approx \frac{2R^{2l-2}}{(l+1)^2}\int\zeta \big\lvert \frac{\delta \rho}{\rho}\big\rvert d^3 x\left(\int\limits_0^R \rho r^{2l+2}dr\right)^{-1}$, where $\zeta = 6.0\times 10^{-59} \left(\frac{3}{2\omega}\right)^2 \rho^2 T^6$ is the bulk viscosity factor. In~principle all these timescales are sensitive to EOS due to the occurrence of density $\rho$ in the equations, }}~\citep{Andersson2001, Haskell2015}:
\begin{equation}
\tau_{GW} = -47 \left( \frac{M}{1.4\textrm{ M}_{\odot}} \right)^{-1} \left( \frac{R}{10\textrm{ km}} \right)^{-4} \left( \frac{P}{1\textrm{ ms}} \right)^6 \textrm{s},
\end{equation}

while the characteristic damping timescales associated with the bulk viscosity, $\tau_{bv}$, and~shear viscosity, $\tau_{sv}$ are
\begin{equation}
\tau_{bv} = 2.7\times 10^{17} \left( \frac{M}{1.4\textrm{ M}_{\odot}} \right) \left( \frac{R}{10\textrm{ km}} \right)^{-1} \left( \frac{P}{10^{-3}\textrm{ s}} \right)^2 \left(\frac{T}{10^8 \textrm{ K}} \right)^{-6} \textrm{s},
\end{equation}
\begin{equation}
\tau_{sv} = 2.2\times 10^5 \left( \frac{M}{1.4\textrm{ M}_{\odot}} \right)^{-1} \left( \frac{R}{10\textrm{ km}} \right)^{5}  \left(\frac{T}{10^8 \textrm{ K}} \right)^{2} \textrm{s},
\end{equation}
where $T$ is the NS core temperature. In~this minimal model it is assumed that at high temperatures, bulk viscosity due to modified URCA reactions provides the main damping mechanism,
while at low temperatures the main contribution is from shear viscosity, due to electron-electron scattering processes. Instability curve from Figure~\ref{fig:rmodesinsta} can be calculated with the following simple formula:
\begin{equation}
\frac{1}{\tau_{GW}}+\frac{1}{\tau_{diss}} = 0,
\end{equation}
where ${1}/{\tau_{diss}} = {1}/{\tau_{bv}} + {1}/{\tau_{sv}} + \text{\em additional processes}$. Several additional mechanisms were considered in the literature, for~example the crust/core velocity difference~\citep{Levin2001, Glampedakis2006}, exotic particles in the core \citep{Andersson2010, Alford2012}, or~strong superfluid mutual friction \citep{Andersson2006, Haskell2014}. These models can lead to very significant changes in the shape, width and depth of the instability~window.

Another natural and astrophysically motivated targets are the binary systems in which the accretion is present. CGW strain from the r-modes excited in accreting systems can be estimated as~\citep{Owen2010, Chugunov2019}:
\begin{equation}
h_0 \gtrsim 1.5 \times 10^{-25} \left(\frac{R}{10 \textrm{ km}} \right)^6  \left(\frac{T}{10^8 \textrm{ K}} \right)^2  \left(\frac{\nu}{600\textrm{ Hz}} \right)^{-1/2}  \left(\frac{M}{1.4\textrm{ M}_{\odot}} \right)^{-1/2}  \left(\frac{d}{1\textrm{ kpc}} \right)^{-1}.
\end{equation}

GWs emitted by a NS destabilised by r-modes carry the information about the internal structure and physical phenomena  {inside} the star and are, in~principle, a~very powerful tool for testing extreme conditions of the NSs interiors, providing information that cannot be obtained by other means. Constraints on the EOS were considered by e.g.,~\citet{Mytidis2015}; the superfluid layer and crust-core interface properties by~\citet{Haskell2015}; rotational frequency evolution in accreting binaries by~\citet{Andersson1999}, spin-down of the young NSs by~\citet{Alford2014}. 

Similarly to Section~\ref{sect:eladeform}, one can define the spin-down limit for r-modes, obtained by assuming that all of the observed change in spin frequency is due to  {the} GWs emission~\citep{Owen2010}:
\begin{equation}
h_{sd} = \frac{1}{d}\sqrt{\frac{45 I_3 \dot{P}}{8 P}} \simeq 1.6 \times 10^{-24} \left( \frac{1\textrm{ kpc}}{d} \right) \left( \frac{\mid\dot{f}_{GW}\mid}{10^{-10}\textrm{ Hz/s}} \right)^{1/2} \left( \frac{100\textrm{ Hz}}{f_{GW}} \right)^{1/2},
\end{equation}
and the corresponding r-mode amplitude parameter~\citep{Lindblom1998}:
\begin{equation}
\alpha_{sd}\simeq 0.033 \left( \frac{100\textrm{ Hz}}{f_{GW}} \right)^{7/2} \left( \frac{\mid\dot{f}_{GW}\mid}{10^{-10}\textrm{ Hz/s}} \right)^{1/2}. 
\label{eq:rmodeamp}
\end{equation}

From the point of view of detecting the r-modes, several methods were used so far to look for their signatures in the LVC data. For~example, the~$\mathcal{F}$-statistic method described in Section~\ref{sect:methods} is so general that can be applied also for r-modes CGWs searches, e.g.,~during supernov{\ae} remnant searches~\citep{Abbott2019e}. No~signal was found, but~interesting fiducial r-mode amplitudes upper limits ($\alpha_{sd} \approx 3 \times 10^{-8}$) were~set. 

Searches for r-modes from pulsars with known sky position require at least three free parameters $f_{GW}, \dot{f}_{GW}, \ddot{f}_{GW}$. Fortunately, for~some pulsars these parameters are well measured. As~shown in~\citet{Caride2019}, the~selection of appropriate ranges of frequencies and spin-down parameters is crucial. In~such a case, for~most pulsars from ATNF Pulsar Database, number of required GWs templates is $\sim10^9$--$10 ^{12}$, comparable in terms of the computational cost to the CGWs searches performed in the past in the O1 and O2 observational runs for triaxial ellipsoid models (see Section~\ref{sect:eladeform}). 

Additionally, a~special type of oscillations may come from the newborn NSs. Following the gravitational collapse, the~proto–neutron star (PNS) radiates its binding energy (about 0.1 M$_{\odot}$) via neutrino emission in a timescale of tens of seconds. A~small fraction of this energy can be released through violent oscillations leading to GWs emission. As~shown in~\citet{Ferrari2003}, initial amplitude of such oscillations for the Galactic PNSs should be larger than $h_0 \gtrsim 10^{-22}$ that it is detectable by the LVC, what is marginally consistent with numerical studies of the axisymmetric collapse of the core of a massive star~\citep{Dimmelmeier2002}. However, such a signal is expected to last for a very brief moment (tens of seconds) and its waveform will depend on multiple variables, such as the poorly known high-temperature dense-matter EOS, resulting with a computationally expensive and uncertain data analysis results. Additionally, Galactic supernov{\ae} events are rare (approximately one per century), making the PNSs GWs unlikely in the LVC~data.

It is worth metioning that r-modes are not the only oscillation type that may, in~principle, lead to the CGWs emission. Initially, the~Chandrasekhar-Friedman-Schutz instability was considered a source of the fundamental fluid mode: the f-mode. However, to~produce CGWs from the f-mode instability, very fast rotational frequency is needed, around $95\%$ of the Keplerian frequency~\citep{Friedman1983,Ipser1991},  {as well as hot, $T\gtrsim 5\times 10^8$ K in accordance with recent models
for the observed cooling of Cassiopeia A~\citep{Page2011,Shternin2011}, non-superfluid matter~\citep{Ipser1991,Passamonti2013}}. Such a situation could be possible only if superfluid core is not formed yet, which would correspond to rapidly spinning, newborn NSs. Nevertheless, further works by~\citet{Gaertig2011,Doneva2013} unveil that relativistic, massive ($M\approx2$ M$_{\odot}$) NSs with realistic EOS can support a wider instability window than  {it} was initially thought, especially for the $l=m=4$ multipole. Another consequence of that result is crucial in the context of binary NS mergers: components of binary NS coalescence and also final product of the merger heat up. If~the remnant is supermassive and fast-spinning (close to its Keplerian frequency) unstable NSs, it may emit post-merger CGW signal. For~the reviews on NSs oscillations types, see~\citet{Kokkotas1997, Glampedakis2018}.

CGWs from oscillations will carry asteroseismological information about NSs interior: its hydrodynamics, superfluidity properties, viscosity and temperature.  {They may indirectly} allow for a distinction between NSs and SSs. Detection of such signals will be invaluable in understanding of the details of newborn and accreting~NSs.

\section{Free~Precession}
\label{sect:preces}

Historically, freely precessing NSs were first considered  {as} good sources of CGWs emission by~\citet{Zimmermann1978,Alpar1985}. Let us adopt a free precessing NS model consisting of a shell containing a liquid. For~a perfectly elastic shell and the inviscid fluid, there will be no energy dissipation: such star will precess with a constant angular velocity and wobble~angle. 

However, a~realistic model requires realistic elasticity and fluidity description, as~well as the internal energy dissipation and external, astrophysical torque mechanisms balancing energy dissipation. Internal dissipation means here that energy losses inside NS are converting the mechanical energy into heat. 
 {Strong interactions between vortices and the normal component (mutual friction, mentioned already in Section~\ref{sect:oscill}), are efficient dissipative channels that can dumped precession~\citep{Shaham1977,Sedrakian1999,Haskell2018b}. Reasoning for a mutual friction in the case of type-I superconducting protons is difficult because of lack of a model-independent predictions for the domain structure and size of type-I superconductor~\citep{Haskell2018b}. However, as~it was shown in~\citet{Wasserman2003}, if~the magnetic stresses are large enough, precession is inevitable. Such enormous magnetic stresses can arise if the core is a type-II superconductor or from toroidal fields $\sim10^{14}$ G, if~the core is a normal conductor~\citep{Wasserman2003}. Observational evidence of free precession  can put strong constraints for the mutual friction and confirm weak coupling of the superfluid to the normal component.}

\begin{figure}
	\centering
	\includegraphics[width=0.35\textwidth]{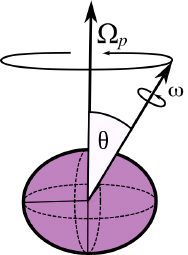}
    \caption{Schematic plot of free precession: rotational axis of the object moves around the precession axis, with~the wobble angle $\theta$ and precession spin $\Omega_p$.}
    \label{fig:precess}
\end{figure}

 {Additionally, if~the magnetic and rotational axes are misaligned, NS will precess~\citep{Mestel1972}. That~process can be especially important in the early life of the millisecond magnetars~\citep{Lander2017,Lander2018}.
For the very young NSs, before~the crust solidifies, any elastic component in the moment of inertia tensor is allowed and effective ellipticity\footnote{In the case of precession, general expression for the ellipticity is defined as $\epsilon = {(I_{3}-I_{1})}/{I_0}$, where moment-of-inertia tensor is given by $I_{ij}=\int \rho(r^2 \delta_{ij} - x_i x_j)dV$ and $I_0 = ({8\pi}/{3})\int \rho_0 r^4 dr$ is the moment of inertia of the spherically symmetric density field $\rho_0$.} $\epsilon_B$ (introduced in Section~\ref{sect:magdeform}, Equation~(\ref{eq:epsilon_B})) comes entirely from the magnetic deformation. When considering precession, $\chi$ from Figure~\ref{fig:btriaxe}, between~axis of rotation and $I_{3}$ becomes the angle around which the rotational axis moves around. It is commonly called the wobble angle and denoted by $\theta$ in Figure~\ref{fig:precess}. Rotational frequency of the slow precession $\Omega_p$ (assuming rigid rotation of the uniform-density object) can be expressed as~\citep{Mestel1972}:}

 {
\begin{equation}
\Omega_p = \omega \epsilon_B \cos(\theta),
\end{equation}
where $\omega=2\pi\nu$. For~non-rigid rotation, the~centrifugal force can deform a star by itself; a quantity corresponding to size of a centrifugal distortion is $\epsilon_{\alpha}$ and both deformation components are related to the global parameters of the object~\citep{Lander2017,Lander2018}:
\begin{equation}
\epsilon_{\alpha} \sim \frac{\omega^2 R^3}{GM} \sim 0.21 \left( \frac{R}{10 \textrm{ km}} \right)^3 \left( \frac{\nu}{1\textrm{ kHz}} \right)^2  \left( \frac{M}{1\textrm{ M}_{\odot}} \right)^{-1},
\end{equation}
\begin{equation}
\epsilon_{B} \sim \frac{B^2 R^4}{GM^2} \sim 1.9\times 10^{-6} \left( \frac{R}{10 \textrm{ km}} \right)^4 \left( \frac{B}{10^{15}\textrm{G}} \right)^2 \left( \frac{M}{1\textrm{ M}_{\odot}} \right)^{-2}.
\end{equation} 
}

 {
Dynamical evolution of the magnetic field in the newborn NSs is crucial to understand their GWs emission. Interplay between energy loss due to the inclination angle evolution $\dot{\theta}$ and the internal energy-dissipation rate due to viscous processes (see Section~\ref{sect:magdeform}) may lead to the  precession damping and change of angle $\theta$. As~it can be deduced from Equation~(\ref{eq:h0magn}), maximal CGWs emission can be expected when rotational and magnetic axes are orthogonal to each other.~Ref. \citet{Lander2018} shows that below $10^{14}$ G, all NSs at some point of evolution have orthogonal rotational and magnetic field axes, regardless of their birth spin frequency. Above~$10^{14}$ G only those NSs that are born spinning fast enough can enter the orthogonalisation region.}
Additionally, energy and angular momentum are carried away as GWs from a freely precessing NSs~\citep{Bertotti1973}. If~the body is rigid enough, $\theta$ diminishes monotonically, but~elastic behaviour may instead increase $\theta$ to $\pi/2$.

While dissipation mechanisms tend to decrease $\theta$, pumping mechanisms, such as e.g.,~accretion, are increasing $\theta$. As~shown in~\citet{Lamb1975} accretion torque can be effective in exciting large amplitude precession of a rigid body with the excitation timescale ($\tau_{ACC}$, introduced in Section~\ref{sect:magdeform}) comparable to the spin-up timescale. The~criterion for excitation is $\tau_{excitation} < \tau_{damping}$. The~evolution of the wobble angle can be therefore expressed as
\begin{equation}
\dot{\theta} = \frac{\theta}{\tau_{excitation}} -\frac{\theta}{\tau_{damping}},
\end{equation}
where in the simplest model one can assume $\tau_{excitation}=\tau_{ACC}$ and $\tau_{damping}=\tau_{CP}$. In~principle, inequality $\tau_{excitation} < \tau_{damping}$ can be used to set upper limits for the GWs amplitudes (for the NSs with known or assumed rotational frequency, accretion rate and crustal breaking strain). Unfortunately, such a signal will be a few orders of magnitude too weak to be detected by the LVC and the Einstein Telescope~\citep{Jones2002}.

GW strain for free precession is given by~\citet{Zimmermann1978,Jones2002,Broeck2005}:
\begin{equation}
h_0=\frac{G}{c^4}\omega^2 \Delta I_d \theta \frac{1}{d} \sim 10^{-27} \left( \frac{\theta}{0.1 \textrm{ rad}} \right) \left( \frac{1 \textrm{ kpc}}{d} \right) \left( \frac{\nu}{500 \textrm{ Hz}} \right),
\end{equation}
where $\Delta I_d = I_3 - I_1$ is the strain-induced distortion of the whole star, also called effective oblateness. Wobble angle $\theta$ cannot be too big, because~too big a deformation of the object (too large breaking strain $\sigma_{max}$ introduced in Section~\ref{sect:eladeform}) may destroy NS, so $\theta_{max} \sim \sigma_{max}$.
Generally, the~CGWs emission originating in free precession process is  present at frequencies  $f_{GW,1}=2\nu$, $f_{GW,2} = \nu + \nu_{p}$, where $\nu_{p}$ is the precession frequency~\citep{Zimmermann1978}; after including second order expansion ($\sim\theta^2$) also $f_{GW,3} = 2(\nu + \nu_{p})$, according to~\citet{Broeck2005}. This third frequency will be seen in GWs spectra with $h_0$ almost two orders of magnitude smaller than first and second frequencies. According to~\citet{Broeck2005}, $f_{GW,1}$ should be detectable if NS spin-down age is much less than $10^3$ yr and observability of $f_{GW,2}$ and $f_{GW,3}$ depends on the crustal breaking strain, which is very uncertain parameter, see Section~\ref{sect:eladeform}. To~extract information about the wobble angle $\theta$ and the deformation $\epsilon$ at least two of the above-mentioned, characteristic frequencies need to be~detected.

Free precession may be much longer lived ($\sim10^5$ years) than initially thought~\citep{Cutler2000}, resulting with weak CGWs emission $h_0\sim10^{-28}$ or smaller. According to~\citet{Jones2002}, it is impossible to find astrophysical pumping mechanisms capable of  {producing} CGWs detectable by an Advanced~LIGO. 

 {Some hints about the precession were delivered by EM observations of the radio pulsars. The~most convincing evidence for free precession is provided by 13 years observations of PSR B1828-11. Variations of arrival-time residuals from PSR B1828-11 were interpreted as a precession, with~precession period $\approx250$ days~\citep{Stairs2000,Akgun2006}. Results obtained from observational data are consistent with the model of precession of a triaxial rigid body, with~a slight statistical preference for a prolate NS shape~\citep{Akgun2006}. Also~analysis of the timing data from another source, PSR B1642-03, also were interpreted as a precessing NS, with~characteristic variation from 3 to 7 years~\citep{Cordes1993,Shabanova2001}.}

Free precession may cause deformations of the object, subsequently allowing testing the NS dense-matter properties, such as breaking strain, viscosity, rigidity and elasticity. CGW detection from free precession will allow also for better understanding of processes present inside the NSs, such as heating, superfluidity, dissipation, mechanisms of the torque and its~evolution.

\section{Summary}
\label{sect:summary}
 {In} recent years we have witnessed the first direct detections of gravitational waves~\citep{Abbott2016b,Abbott2017a,Abbott2017b, Abbott2017c, Abbott2017d,Abbott2018b}. It~started a new observational field: the gravitational wave astronomy.  {Those} detections  {corresponded to observation of} coalescences of  {the} binary systems: violent and energetic phenomena involving black-holes, impossible to detect with traditional astronomical methods, as~well as electromagnetically bright binary neutron stars system coalescence. As~the interferometers improve their sensitivity~\citep{Harry2010,Acernese2014,MooreCB2015}, we expect other types of signals to be detected. One promising scenario is the detection of continuously emitted, periodic and almost-monochromatic GWs produced by rotating NSs. Several mechanisms can be responsible for such a GWs emission: elastically and/or magnetically driven deformations (mountains on the NS surface supported by the elastic strain or magnetic field), thermal asymmetries due to accretion, free precession or instabilities leading to modes of oscillation (r-modes).

Discovery of a persistent source will be the capstone of GW astronomy will allow for repeatable  {observations, which will result in acquiring more knowledge and better understanding of the} NSs interiors, especially on their crust physics and elastic properties at low temperatures (study of cold EOS as opposed to hot EOS in binary NS collisions). In~general, these NSs may have different masses and matter conditions than the objects in binary NS inspirals and mergers, which leads to valuable GWs observations in different dynamical regimes and, additionally, to~various tests of the general relativity by exploiting the fact that detectors are moving with respect to the source (e.g., studies of independent wave polarizations).  {In principle, NSs observed in CGWs could be also a valuable tool in the detectors' strain calibration, astrophysical distance ladder calibration (useful in cosmology) and measurements of fundamental aspects of gravity~\citep{Pitkin2016,Isi2017}. }  

The most promising CGWs emission scenario (giving the highest GW amplitude)
 assumes deformations on the NS surface. Because~of that, most of the LVC effort is concentrated on searches of signals consistent with the model of triaxial ellipsoid radiating CGWs at twice the spin frequency~\citep{Abbott2017k, Abbott2018c, Abbott2019b, Abbott2017c}. However, in~the future, when LIGO and Virgo detectors will reach their design sensitivity and new instruments (ground- and space-based,~\citealt{Akatsu2017,Amaro2017,Sathyaprakash2012,Unnikrishnan2013}) will join the network, testing of other models and CGWs sources will be~possible. 

 {Additionally, by~combining CGWs and EM observations one can break the degeneracy in some expressions considered in this paper. For~example, for~the triaxial ellipsoid model, by~knowing distance and spin frequency from EM observations and CGW amplitude from GWs detections, immediately one can obtain size of the deformation. This quantity is a few order of magnitudes different for NSs and SSs and carries information about the crust properties. For~the strongly magnetised objects, such as magnetars, newborn NSs or NSs in LMXB systems joint EM and CGWs analysis will allow for testing magnetic field composition, strength and evolution. Also unstable r-modes carry information about the NSs interiors: their superfluid layer, viscosity and temperature. Finally, the~precession---in the case of detection may expose information about magnetic field, elasticity and superfluidity of the~NSs. }

Potential CGWs detection will allow for testing NSs matter (their EOS, crustal properties, deformability), processes inside the object (heating, superfluidity, instabilities, magnetic field evolution), close environment of the stars (their magnetic field, accretion processes). Such unique analysis will be essential also in our understanding of NSs evolution, especially for the newborn objects. Very young NSs undergo several dynamical mechanisms and usually are unstable - such changes are difficult to observe in traditional, EM telescopes. GW astronomy will open a new door for NSs studies and may be a crucial piece in solving  {the} mystery of NSs nature and~EOS.   
\vspace{6pt}

\section*{Funding}
The work was partially supported by the Polish National Science Centre grants no. 2016/22/E/ST9/00037, 2017/26/M/ST9/00978 and 2018/28/T/ST9/00458.

\section*{Acknowledgements}
The authors thank Brynmor Haskell, Karl Wette, David Keitel for useful insights and comments that greatly improved the manuscript and Maria Alessandra Papa for suggesting useful references. The~authors thank Damian Kwiatkowski for help with language~editing.



\end{document}